\documentclass{article}

\usepackage{arxiv}
\usepackage{titlesec}
\usepackage{titling}

\usepackage{changepage}
\usepackage{amsthm}
\usepackage[utf8]{inputenc}
\usepackage[T1]{fontenc}    
\usepackage{hyperref}      
\usepackage{url}          
\usepackage{booktabs}      
\usepackage{amsfonts}       
\usepackage{nicefrac}      
\usepackage{microtype}    
\usepackage{lipsum}
\usepackage{graphicx}
\graphicspath{ {./images/} }
\usepackage{fancyhdr}
\usepackage{parskip}

\titleformat{\section}
  {\normalfont\fontsize{12}{15}\selectfont}{\thesection}{1em}{}

\begin{document}

\title{The Decision Path to Control AI Risks Completely: \\ Fundamental Control Mechanisms for AI Governance}
\author{Yong Tao\thanks{Fellow, Society of Decision Professionals (SDP); PhD, Decision and Risk Analysis, Stanford University} \\ (In memoriam: Ronald A. Howard\thanks{Professor Emeritus, Department of Management Science and Engineering, Stanford University} : 1934-2024) }
\date{\today}

\maketitle

\begin{abstract}
Artificial intelligence (AI) advances rapidly, but achieving complete human control over AI risks remains an unsolved problem, akin to driving the fast AI "train" without a "brake system." By exploring fundamental control mechanisms at key elements of AI decisions, this paper develops a systematic solution to thoroughly control AI risks, providing an architecture for AI governance and legislation with five pillars supported by six control mechanisms as building blocks, illustrated through a minimum set of AI Mandates (AIMs). Three of the fundamental "how-to" control mechanisms need to be built inside AI systems and three to be established in society to address the broad spectrum of major areas of AI risks: 1) align AI values with human users; 2) constrain AI decision-actions by societal ethics, laws, and regulations; 3) build in human intervention options for emergencies and shut-off switches for existential threats; 4) limit AI access to resources to reinforce controls inside AI; 5) mitigate spillover risks like job loss from AI. We also highlight the differences when these control mechanisms are applied to physical AI systems versus generative AI models. In addition, we discuss how to strengthen analog physical safeguards to prevent smarter AI/AGI/ASI from circumventing core safety controls by exploiting AI's intrinsic disconnect from the analog physical world: AI's nature as pure software code run on chips controlled by humans, and the prerequisite that all AI-driven physical actions must be digitized. These findings establish a theoretical foundation for AI governance and legislation as the basic structure of a "brake system" for AI decisions. If implemented, fundamental control mechanisms can rein in AI dangers as completely as humanly possible, removing large chunks of currently wide-open AI risks, substantially reducing overall AI risks to residual human errors. The proposed framework allows society to manage AI threats effectively and efficiently, safeguarding humanity by controlling the most threatening AI risks deep inside AI decision processes, reinforced by a robust societal AI governance infrastructure.
\end{abstract}


\renewcommand{\thefootnote}{\fnsymbol{footnote}}

\section{Introduction}
\paragraph{}The rapid growth of artificial intelligence (AI) capabilities presents significant potential benefits, but also considerable harms and catastrophic risks \cite{yoshua_bengio_53350e8f}, \cite{holden_karnofsky_e480d4ed}, for which we currently lack effective control mechanisms. Consequently, we are advancing into a future shaped by AI, an uncharted territory filled with uncertainties and risks.
\paragraph{}In this paper, we use "AI" to refer broadly to generative AI models, AI agents based on large language models (LLM), small language models (SLM), and physical AI systems that perform independent decision-making tasks. These AI decisions can indirectly influence or directly drive physical actions, depending on the context.
\paragraph{}There is a general consensus that existing laws and regulations are inadequate to address the risks AI poses to humans. Although some developers have proposed and implemented specific “guardrails” for certain frontier AI models voluntarily, and the implementation of the EU AI Act is underway, most of the current AI governance research remains philosophical and aspirational, lacking systematic, concrete legislative actions that are urgently needed. In addition, the intense race to achieve artificial general intelligence (AGI) increases the risks of economic instability, societal disruption, and existential threats. The absence of effective governance is a significant risk in itself, similar to driving the AI "train" without a functional “brake system,” leaving substantial AI risks unaddressed.
\paragraph{}Each action taken by AI stems from a decision made by AI. In contrast to conventional laws, legislation concerning AI must govern not only the AI itself but also its human developers, users, and even individuals who interact with or are impacted by AI's decisions. Consequently, extensive interdisciplinary studies are crucial, which require a blend of knowledge in the fields of AI technology, ethics, laws, and decision sciences to create a solid framework for AI governance and legislation. This shift from fragmented high-level discussions to structured, decision-based AI legislation signifies a new approach to protecting humanity from both known and unforeseen AI threats.
\paragraph{}To facilitate a thorough exploration, we must step back and broaden our perspectives with a few fundamental understandings of AI. We begin by defining what AI governance entails.
\subsection{AI Governance}
\paragraph{}AI governance is the collective efforts of AI developers and deployers, users, legislators, and the public to maximize its societal benefits and minimize potential risks from AI at the same time. While AI benefits are driven by market force, the risk management (RM) elements in AI governance include three lines of defense (Figure 1): market force and ethics, existing laws, and new AI legislation.
\paragraph{}Clusters of AI risks have been identified, some through accident reports \cite{peter_slattery_99a31fe8}, others through plausible scenario analysis. It is essential to recognize that risk management involves evaluating the trade-off between the benefits of risk mitigation and the costs incurred from these actions. Insufficient and excessive risk management are both suboptimal. Although some risks may be eliminated, it is more common to reduce risks to a level deemed acceptable relative to the cost of mitigation. Therefore, AI risk management should be viewed in terms of acceptable risk thresholds and corresponding costs. Each specific AI risk may be addressed by one line of defense, or the risk may move to the next line. If unacceptable risks persist after all existing defenses, new AI governance mechanisms or legislation is needed.
\subsubsection*{A. First Line of Defense: Market Force and Ethics}
\paragraph{}This line represents the voluntary, yet vulnerable, market mechanism that incentivizes AI developers to create useful, secure, and safe AI models primarily for their customers and users. Some risk mitigation requirements, such as cybersecurity, have become key design features of AI products due to market force. When conflicts arise, the ethical practices of AI developers come into play to influence product design decisions. Most large AI developers have established internal AI ethics boards \cite{jonas_schuett_353a9c62}.

\begin{figure}[ht]
\centering
\includegraphics[width=0.75\textwidth]{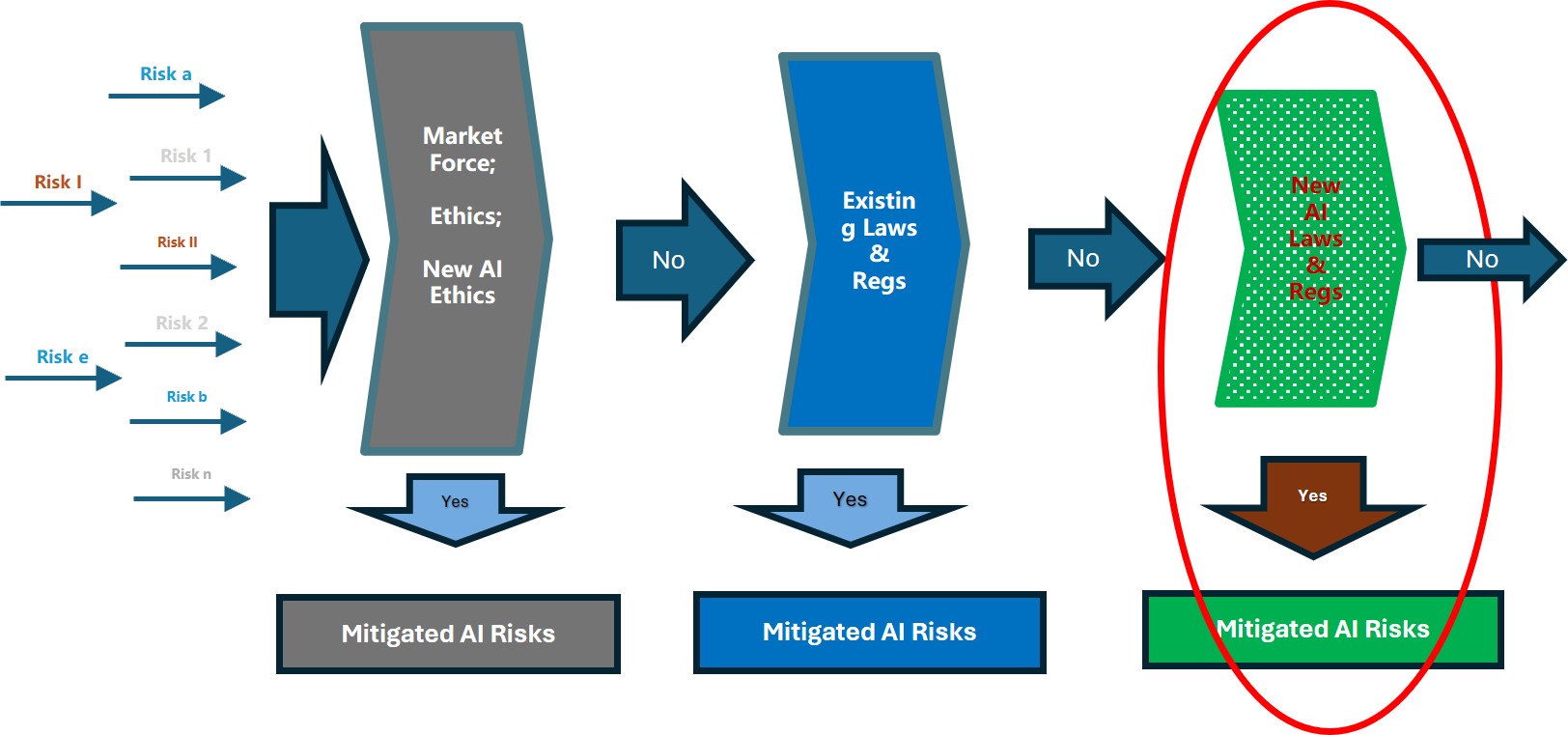}
\caption{Lines of Defense Against AI Risks}
\label{fig:figure1}
\end{figure}

\paragraph{}Ethics covers a wide range of issues, but is not legally binding; it reflects the voluntary goodwill of AI developers and engineers. Consequently, ethical practices can vary widely among developers and engineers. This line of defense is effective in mitigating AI risks when it aligns with the interests of both developers and their customers, but less so when it does not.
\paragraph{}So far, many AI risks have been reasonably well managed through market force and ethics. Cybersecurity risks in AI systems have been effectively addressed through market force that provides strong incentives—an AI system with inadequate cybersecurity is unlikely to attract many users. The focus of this paper is on the risks that market force and ethics do not cover well.
\paragraph{}User-friendly interfaces for soliciting and representing user values and preferences by AI systems have also been reasonably well achieved through market incentives, although there is room for improvement as AI systems become more integrated into the human-AI decision hierarchy to assist with major human decisions where user preferences may be subjective, complex, uncertain, and dynamic. AI systems that do not effectively solicit user preferences through friendly interfaces are less likely to be adopted.
\subsubsection*{B. Second Line of Defense: Existing Relevant Societal Laws and Regulations}
\paragraph{}If a risk penetrates the defense of market force and ethics, it then encounters existing societal laws and regulations \cite{hilke_schellmann_ac025307}. Copyright and privacy lawsuits filed by artists against AI developers for allegedly training models without permission illustrate this line of defense. Many existing laws need to be expanded to address AI. Most of the 18 AI laws passed in California in 2024 primarily expand coverage from humans to AI.
\paragraph{}However, applying existing laws to AI systems presents inherent challenges due to the differences between human and machine intelligence.

\subsubsection*{C. Third Line of Defense: New AI Governance Legislation}
\paragraph{}If a risk penetrates the first two lines of defense, it becomes necessary for society to establish new AI laws and regulations to mitigate that risk. This type of AI risks keeps many experts, legislators, users, and the public awake at night. This paper does not intend to propose laws to deal with such specific AI risks in a piecemeal fashion. We aim for a systematic approach to develop a comprehensive AI governance legislation that eliminates and reduces all major areas of unaddressed AI risks. This new AI governance legislation should be limited to a minimum of essentials, not to stifle AI innovation. The new AI legislation is part of AI governance, in addition to ethics and existing laws and regulations. The fundamental control mechanisms that we explore in this paper are intended to be part of the new AI legislation. Of course, these control mechanisms can be practiced as ethics by AI developers before they are enacted into law.
\paragraph{}Examples of AI risks that warrant new laws include existential threats from AI systems, widespread disruptions to social order caused by generative AI, and challenges posed by self-driving cars and humanoid robots roaming the streets soon. We must also set limits on the freedom of operations of AI systems by restricting AI access to resources to ensure that we reap the benefits while limiting potential harm to society.
\paragraph{}There will also be challenges in assessing accountability under existing laws when AI systems are involved in accidents that cause harm, as well as issues of bias, discrimination, and privacy violations. Many existing laws and regulations need to be expanded to specifically address AI use cases.
\paragraph{}Currently, the most significant gap in AI governance lies in the third line of defense: the lack of new systematic AI governance legislation, the primary focus of this paper. The need for systematic and concrete legislation is particularly urgent for countries leading the development of AI, such as the US and China.
\paragraph{}Although the EU AI Act, a landmark piece of legislation, was enacted in November 2023 with a "block and verify" approach\cite{mag__junaid_sattar_butt_d9057b15}, concrete legally binding control mechanisms are lacking and are still being developed. The findings in this paper can greatly enhance the implementation of the EU AI Act, which is phased over several years \cite{holger_hermanns_fdf7bc88} to gradually incorporate concrete specifics, such as the recent Code of Practice \cite{lily_stelling_6bc49d11}.
\paragraph{}Although many studies on AI governance focus on corporations and organizations, this paper addresses the need for systematic AI governance and legislation at the country or state level. We believe that governance at this level is essential to defend against all AI risks and provides the foundation for the management of AI risks in corporate and organizational settings.

\subsection{AI Governance Goals}
\paragraph{}Although there is a lack of systematic concrete legislation currently, given the imperative to manage AI in society, the general consensus on AI governance goals converges into four main areas:
\begin{enumerate}
\item Harnessing the benefits of AI for users and society
\item Establishing boundaries for AI decisions and actions to prevent social chaos
\item Maintaining human control in AI-triggered emergencies; and
\item Mitigating spill-over risks that AI causes, such as job loss.
\end{enumerate}
\paragraph{}While prominent AI developers focus on advancing technology to harness its benefits, governments and societies are tasked with regulating AI to address its current and future risks. This separation of duties may result in an imbalance, where innovation takes precedence over essential safety, cybersecurity, and ethical concerns, particularly due to the complexity of AI technologies.
\paragraph{}An effective and efficient AI governance system would likely arise from a combination of incentives and legislation, along with a potentially specialized AI governance authority. This authority, empowered by a broad range of AI stakeholders and the public, could provide adequate guidance to AI developers and users, facilitating the rapid adoption of innovative AI products and services while simultaneously safeguarding human society. The participation of various stakeholders can ensure that governance mechanisms are practical and widely accepted, fostering a delicate balance between innovation and public safety.
\paragraph{}AI governance goals serve as essential guides for AI development and legislation. Many declarations from global AI summits clarify these desires and needs at the highest level and with the broadest global participation. Most AI developers currently conduct their own safety and security research as a form of voluntary governance over their AI system capabilities; such efforts are necessary to explore technical solutions and establish a knowledge base for systematic AI legislation. However, these efforts are not sufficiently neutral, independent, or authoritative to earn public trust.
\paragraph{}To prevent conflicts of interest, AI governance legislation, like any law, should be enacted and enforced primarily by individuals who are not closely involved in AI development. Society and its legislators are tasked with enacting AI laws to protect everyone. This separation of powers is crucial to ensure that AI governance prioritizes public welfare over the interests of specific AI developers and users, but the complexities of AI technology require deep participation from AI developers and technology experts.

\subsection{The Urgent Challenge: Finding a Sound Solution to Control AI Risks}
\paragraph{}Rapid advances in AI, coupled with the lack of robust governance frameworks, have raised public concern. According to the Stanford AI Index Report \cite{nestor_maslej_e97e0966}, \cite{maslej2025artificialintelligenceindexreport}, more than half of Americans expressed greater concerns than excitement about AI \cite{colleen_mcclain_feab2312}. As AI capabilities grow, so does awareness of its potential dangers. As the proverb suggests, "you can't lift the stone without being ready for the snake that's revealed." The public perceives inadequate protection against the potential risks associated with AI \cite{andeed_ma_f2ac3d19}.
\paragraph{}The current AI regulatory landscape \cite{weiyue_wu_63b8028b} remains primitive compared to the speed of AI advancements. In response to significant public incidents and accumulated risk-based research, the EU AI Act was passed in November 2023. This legislation adopts a risk-based approach to manage the high and limited risks posed by AI to humans while eliminating unacceptable-risk AI and permitting minimal-risk AI \cite{oskar_josef_gstrein_de9f9e7e}. Although this may alleviate concerns that regulations are constantly lag behind technological developments, complex questions about its effectiveness and enforceability persist due to the absence of fundamental control mechanisms for AI systems and developers, as well as guidance for users and law enforcement. A purely risk-based regulation is a good starting point, but insufficient to prevent catastrophic risks and existential threats, leaving dangerous loopholes open.
\paragraph{}Despite these concerns, the EU AI Act is a significant step in regulating AI \cite{maurizio_bottini_1ee693d7}. It functions as a flexible barrier to manage AI risks, relying on subjective judgments from EU AI Office officials and representatives. This makes it a unique experiment in AI governance legislation.
\paragraph{}To overcome the limitations of the risk-based approach, several value-based studies aim to align AI values with human values \cite{stuart_russell_857f83a4}. Perhaps infuse positive moral ideas like peace and motherly love into AI if possible. Although this is a positive step, it remains inadequate for controlling risks of AI systems, which often operate as "black boxes" with decisions that are not fully explainable. Value-based AI research provides a proactive strategy for guiding AI development, but translating its complicated soft guidance, such as factoring human values into giant utility functions without distinguishing user preferences and general human values such as ethics, laws, and regulations, can be extremely challenging.
\paragraph{}Neither the US nor China has established comprehensive national AI governance legislation. Although AI-related legislation in the US has increased rapidly in recent years \cite{vikram_kulothungan_e0a67a51}, it remains fragmented between various federal and state authorities. China has rapidly created administrative guidelines similar to the EU AI Act, establishing aspirational goals through central authorities overseeing AI development.
\paragraph{}Coherent and comprehensive AI legislation is essential for the long-term sustainability of AI and society. Legislation forms the foundation for social order, increasingly shaped by technological advances. Just as it is prudent to design brakes for a new fast train, AI governance legislation provides the necessary brake system for sustainable AI development. However, current research lags behind the pace of AI advancements, limiting the ability of nations to enact coherent and convincing AI governance legislation \cite{gary_marcus_e91dc1b5}.
\paragraph{}This makes AI governance legislation one of the most urgent challenges of our time. It is imperative for the US and China to enact comprehensive AI governance legislation as leaders in AI development. Although overly restrictive legislation can stifle innovation, the absence of a minimum set of AI laws could lead to unintended consequences for these nations before they can address potential competitive threats. Thus, a balance between AI legislation and development must be maintained.
\paragraph{}In this paper, our aim is to advance AI governance research by exploring the foundational role of decisions in AI legislation. We propose a decision-based paradigm for AI governance, emphasizing fundamental control mechanisms for AI decisions rather than solely relying on risk-based research and value alignment efforts.
\section{Why We Need a Decision Paradigm for AI Governance and Legislation}
\paragraph{}Why is it difficult to grasp AI governance? Why do our familiar methods for governing other technologies fail when applied to AI? What are our greatest fears from AI? Existing AI governance approaches often fall short of addressing these core issues, highlighting the need to step beyond the realm of AI technology itself and examine the broader decision role that AI plays in human society. This shift calls for a decision-based approach to AI governance and legislation \cite{anton_sigfrids_1a758b34}.
\paragraph{}This section outlines the fundamental premises of our research: First, we compare AI with other technologies. Next, we compare the governance of human decision actions with that of AI. Finally, we describe how a decision-based paradigm can serve as the foundation for AI governance.
\subsection{AI as Extensions of the Human Brain}
\paragraph{}Hammers, wrenches, guns, cars, chips, nuclear bombs—all are passive technology tools that extend the capabilities of human hands that enable us to act more efficiently. Decisions on the development and use of such technologies are made by humans, governed by the ethics, laws, and regulations of their specific industries. Developers provide technology products and ensure specific performance, while users apply these technologies responsibly.
\paragraph{}The governance of all these other technologies has focused on the producers and users of technology products \cite{nicholas_emery_xu_81d3b140}. We have governed these technologies by regulating human decisions about their development and use. This approach works because these technologies are passive tools. Ultimately, users bear the consequences of their actions while producers assume liabilities of their technology products.
\paragraph{}AI differs fundamentally because it extends the human brain \cite{ximeng_chen_610e243b}: AI makes decisions that lead to actions that affect humans. Before AI comes into the scene, humans are governed by our decisions and actions. As humans delegate more and more decisions for AI to make independently, chaos will ensue if AI decisions do not adhere to the same ethics, laws, and regulations that govern human decision actions. This requires a different governance approach for AI relative to other technologies, focusing on regulating AI decisions \cite{lan_xue_c9da6caf} and actions. We need to explore an approach in which AI decisions are governed with relevant existing societal values and legal frameworks \cite{mark_anthony_camilleri_82213efb}.
\paragraph{}Mass-use technologies are triple-edged swords: providing potential benefits, harboring risks of accidents, and running the risks of being used by malicious actors. Although we recognize the good and bad intentions behind the use of technologies, we often underestimate the significant impact of accidents. For instance, in 2023 in the US, 40,990 people died in car accidents, surpassing the 36,574 US soldier deaths during the Korean War. Governance legislation for pervasive technologies like AI must address, eliminate, and reduce the effects of accidents and effectively deter malicious uses.
\subsection{Key Similarities and Differences between AI and Humans}
\paragraph{}AI and humans both possess cognitive and independent decision-making abilities. However, fundamental differences set them apart. What are these differences and how can we leverage them for human control over AI?
\paragraph{}As shown in Table 1, the most fundamental difference between AI and humans is that AI is synthesized intelligence embodied in chips that are inherently NOT under its control. AI, AI agents, artificial general intelligence (AGI), and even artificial superintelligence (ASI) are essentially software embedded on electricity-powered chips provided by humans. Although chips can be empty without AI code, AI cannot exist without the material base of these chips. Currently, AI engineers must meticulously manage AI hardware and chips to ensure AI functionality. AI cannot control its own destiny because it lacks innate control over the hardware that forms its material foundation. This presents both strengths and weaknesses for AI, contrasting with the human brain which is inseparable from the human body. The separability of AI software and hardware allows for easier enhancement through scaling, enabling AI to seamlessly connect with other software or AI modules, thus increasing its capabilities much faster than human learning \cite{dan_hendrycks_a6f9ec1b}. Furthermore, AI code can be easily copied onto numerous machines to perform multiple user tasks simultaneously.

\begin{table}[ht]
 \caption{Human-AI Similarities and Differences}
  \centering
  \begin{tabular}{|c|c|c|}
    \hline
    Characteristics  & Humans   & AI/AGI/ASI  \\
    \hline
    Birth of Life & Birth   & Initial Release     \\
    \hline
    End of Life     & Death & Erase All Copies      \\
    \hline
    Material Base     & Carbon       & Silicon  \\
    \hline
    Brain-Body Separability   & Inseparable  & Separable  \\  
    \hline
    Self-Sufficiency     & Yes       & No  \\
    \hline
    Expandability    & No       & Yes  \\
    \hline
    Easily Copyable   & No       & Yes  \\
    \hline
    Connectivity     & Self       & Other AIs  \\
    \hline
  \end{tabular}
  \label{tab:table}
\end{table}

\paragraph{}The separability of AI from and reliance on its underlying hardware of chips is fundamental for human control. Although the information and AI era has witnessed rapid digitization, the natural world remains analog. Energy and electricity are essentially analog as well. If humans can physically manage the power switches for AI chips without completely digitizing that control, effective human control can be maintained. However, the vast number of humans using AI, coupled with their diverse motivations and AI's increasing persuasive power, makes enforcing such requirements and ensuring complete human control extremely challenging. However, this intrinsic separability forms the foundation for human control over AI. We will revisit this feature in later sections.
\subsection{AI Capabilities vs. AI Decisions}
\paragraph{}The ongoing debate surrounding AI governance often conflates AI capabilities with AI decisions, leading to considerable confusion. To effectively address AI governance, it is crucial to differentiate between AI decisions and AI capabilities. AI capabilities represent what AI models can or cannot do. These capabilities currently specialize in particular areas while striving toward artificial general intelligence (AGI), which would excel across all domains of intelligence. In contrast, AI decisions pertain to what AI will do in specific use cases and scenarios, determined by the model, context, and set of user instructions. The same AI capabilities can manifest itself in numerous AI decisions, each potentially having different impacts, similar to the way the human brain makes diverse decisions across various aspects of life. Only during the decision-making process do intelligence capabilities get balanced by the consideration of human values to the degree of decision maker's maturity. 
\paragraph{}We can draw valuable insights from the development of human intelligence and governance. Human cognitive and decision-making capabilities are slowly cultivated over about the first 20 years of life and continue to develop at a slower pace through real-life decision-making and lifelong learning. During this formative period, individuals benefit from the guidance of parents, guardians, teachers, and society, cultivating decision-making skills. We learn through trial and error, feedback, and structured education, accumulating knowledge and human values that enable us to make independent decisions. Feedback from the outcomes of our choices, whether positive or negative, shapes our intelligence and maturity. Although knowledge acquisition slows after formal education, the practice of independent decision-making is just beginning.
\paragraph{}Lessons from human governance reveal that directly governing human intelligence or thoughts is impossible. The thoughts of a person are private, unobservable, and harmless to others if they are not acted on. If an individual refrains from making decisions that affect others, we should leave her and her thoughts alone. Even prisoners are allowed to think freely, with constraints only on their actions. Society has learned that governing human cognition and intelligence capabilities is unnecessary and impractical, since they do not directly impact others. Human governance instead focuses on our decisions and actions that interact with others in society. A 2020 study in Nature Communications by Tseng and Poppenk indicates that humans generate more than 6,000 thoughts a day \cite{julie_tseng_739e1f04}, but only a small fraction leads to decisions that result in actions, as most thoughts are filtered out through ethics, laws, and regulations (ELRs). It is much more practical and effective to govern human decisions and actions rather than their capabilities of intelligence and thoughts.
\paragraph{}Unlike the human growth process, during the development and  training phases measured by months, AI capability growth is focused on intelligence. Of course, this AI intelligence growth phase itself is governed by existing laws on intellectual property, privacy, nondiscrimination, etc. It is not clear how much AI has learned about societal values during this process and whether AI has the maturity to make decisions independently. Here, maturity means the ability to be considerate of others in terms of acting according to ethics, laws, and regulations. Once a trained AI model is released publicly, it begins to make independent judgments and decisions, with its "thoughts" revealed through human-AI interactions, specifically user prompts and model outputs. Many user feedbacks are probably trying to teach AI human values, but currently AI models do not yet learn dynamically from user interactions in a real-time fashion, they only learn through training by specific datasets through "batch" jobs with subsequent model releases. One thing is for sure, currently, AI outputs and decisions are not subject to the same level of enforcement of societal ELRs as humans, which is one major reason why AI lags in maturity in adhering to human values.
\paragraph{}We can apply lessons from human governance to AI governance. AI capabilities are emergent, unobservable, and value-neutral: the same capability can be used for beneficial purposes, harmful actions by malicious actors, or accidents can occur. Classifying static AI capabilities as "good" or "bad" is not meaningful; it depends on how those capabilities are deployed and the decisions made under specific circumstances. In this sense, AI capabilities themselves are not governable. In contrast, AI decisions are specific choices made in a real environment that have ethical, legal, and regulatory implications \cite{danielle_s__allen_9c98e0bd}. Decisions establish the necessary context to distinguish between ethical and unethical, legal and illegal AI actions, and to evaluate whether AI decisions yield benefits or cause harm. This understanding is fundamental to AI governance. Thus, the governance of AI means the governance of AI decisions, just as the governance of humans is realized by governing human decisions and actions.
\paragraph{}Even if we try to govern AI capabilities, efforts to control them merely affect the pace of AI development—whether to stop progress, slow it down, or allow it to proceed unchecked. Such approaches have led to dead ends, resulting in moratoriums on AI development or letting AI run wild \cite{emmanouil_papagiannidis_1fc1e900}. The plea for a moratorium from AI experts and world leaders has had limited impact, primarily raising public awareness of AI risks. Key stakeholders in AI are generally unwilling to stop or slow development, arguing that it is  like "throwing the baby out with the bathwater." History shows that market force dictates the pace of technology and AI advancements, characterized by cycles of boom and bust. Currently, the speed of AI progress is limited only by the availability of skilled talents, training data, compute power, and electricity.
\paragraph{}Ultimately, the impacts of AI on humans are from its decisions that drive digital or physical actions. As AI's capabilities manifest through these decisions and actions, both positive and negative outcomes may arise. By examining the steps of the AI decision process, we can identify the root causes of AI risks, creating opportunities to devise optimal solutions to mitigate those risks. Some solutions may need to be embedded inside AI algorithms, becoming intrinsic to AI's capabilities. An AI system that incorporates effective control solutions into its capabilities can make high-quality decisions consistently adhering to societal ELRs.
\paragraph{}Governing AI capabilities and AI decisions are distinctly different endeavors. Human instincts often lead us to focus on the risks associated with AI's remarkable capabilities, which may result in unnecessary overregulation. Ee focus our AI governance efforts on the dynamic decisions and actions resulting from the exercise of AI capabilities. It marks the point of departure of this paper from other work on AI governance.
\subsection{Decision Is All We Get from AI}
\paragraph{}The primary reason humans develop AI is decision-making. Although AI is powerful in areas such as learning, language, and creativity, these abilities ultimately serve one key goal: enabling AI to make decisions on its own or helping humans make better decisions. At its core, AI is about making decisions. Sound decisions depend not only on the intelligence of the system, but also on how effectively it incorporates human values and applies them during the AI decision process. This is similar to the human brain, where decisions control the translation of thoughts into actions that shape the real world. AI-driven decisions may be simple, such as choosing which song to play, or more complex, such as forming a stance on a major issue (an informational decision), or highly consequential, such as approving a loan or guiding life-saving medical treatment.
\paragraph{}AI developers pursuing artificial general intelligence (AGI) are increasingly prioritizing capability gains over AI governance. From a decision-theoretic perspective, the control frameworks needed for AGI are fundamentally similar to those already employed for today’s advanced AI systems, which in many domains surpass human performance. An AGI system primarily enhances the quality of decisions made by current AI systems while preserving the underlying decision structure. As a result, decision-based AI governance and legislation can be consistently applied across current AI systems, AGI, and even ASI.
\paragraph{}Effective AI governance hinges on the regulation of the decisions, actions, and outcomes of AI systems, similarly to how laws and ethical standards govern human decisions, actions, and outcomes. When AI systems participate and mix with human actors, AI decisions must comply with the same ethical and legal standards that regulate human behavior, ensuring AI systems are aligned with human values and legal systems. 
\paragraph{}The decision paradigm for AI governance treats AI decisions the same way as human decisions. This framework aims to establish a foundation for consistent AI governance legislation and will also explore how it relates to policy suggestions from value alignment initiatives and risk-based strategies. By thoroughly examining the AI decision-making process, we can pinpoint crucial intervention points that lead to successful AI governance solutions. We begin with a high-level review of the process of decision analysis to achieve good decision quality.

\section{The Decision Foundation of AI Governance and Legislation}
\paragraph{}In recent decades, theories about human decision-making have evolved in two directions: normative and descriptive. Normative theories prescribe how people should make decisions, such as decision science and decision analysis \cite{ronald_a__howard_5c4c8e1f}, utility theory, decision quality \cite{carl_spetzler_d7449072}, and game theory. Descriptive theories explain how people actually make decisions, highlighting various biases and fallacies that challenge some of the foundations of normative decision theory. Despite this, when faced with important decisions, humans strive to be “rational” to the extent possible and practical. Decision analysis with the decision quality framework offers a practical decision-making methodology based on utility theory, incorporating debiasing techniques.
\paragraph{}Behavioral economists and psychologists, the main contributors to descriptive theories, have documented various human biases that influence our decision-making \cite{daniel_kahneman_8e66163f}. According to Kahneman, there are two systems for human decision-making: system 1 and system 2. System 1 is intuitive and fast, subject to numerous potential biases; system 2 is analytical and slow, also susceptible to many biases. To mitigate some of the most pervasive biases, psychologists proposed prospect theory, a modified decision methodology that incorporates practical rationality.
\paragraph{}AI decision-making primarily follows the practical normative rational decision approach, seeking solutions to constrained optimal decisions under uncertainty, subject to biases. This is largely because it does not make sense to intentionally introduce biases into AI decisions. Even if AI would attempt this, it would be unclear which direction to bias toward, given the hundreds of documented biases. Of course, AI also seeks to reduce biases, but eliminating them entirely is challenging. AI decision-making functions as an analytical process of system 2 operating at the fast speed of system 1, thanks to the rapidly increasing speed of computation.
\paragraph{}Regarding the AI decision-making process, each AI developer may have slightly different variations depending on their market focus and the architecture of their AI systems. However, AI developers strive to adhere to a practical rational decision-making process that is rigorously supported by decision science, decision analysis, and utility theory. At a high level, this process can be described in six steps that interweave around three key decision elements. 
\subsection{Genesis of a Decision}
\paragraph{}The most fundamental premise of human decision making is the recognition that the decision maker has the ultimate agency of independent power to make his or her decisions. A human decision can originate from many different sources. The moment the decision-maker declares her need to make a decision marks the genesis of that decision. Currently, AI does not possess the proactive capability to independently identify decision-making needs and "declare" the need for a new decision. AI may be able to decompose a decision goal into subgoals. The premise underlying the creation of AI is to place it within a human-AI decision hierarchy, where the genesis of the decision comes from a human, who may choose to delegate her decision to AI, which then makes the decision on her behalf or assists in making it. Humans are responsible for the meta-level decision of whether and to what degree to use AI.
\paragraph{}The genesis of human decisions is closely related to unmet human goals shaped by emotions and conscience, which AI does not currently possess. As AI progresses toward artificial general intelligence (AGI) or artificial superintelligence (ASI), the potential for the genesis of decisions by AGI or ASI may require further examination.
\subsection{Framing} 
\paragraph{}Framing a decision involves identifying the decision maker's perspective, providing context, defining the scope, and establishing the relationships between the decision in focus and past or future decisions. Framing is the first step in the decision process.

\begin{figure}[ht]
\centering
\includegraphics[width=0.75\textwidth]{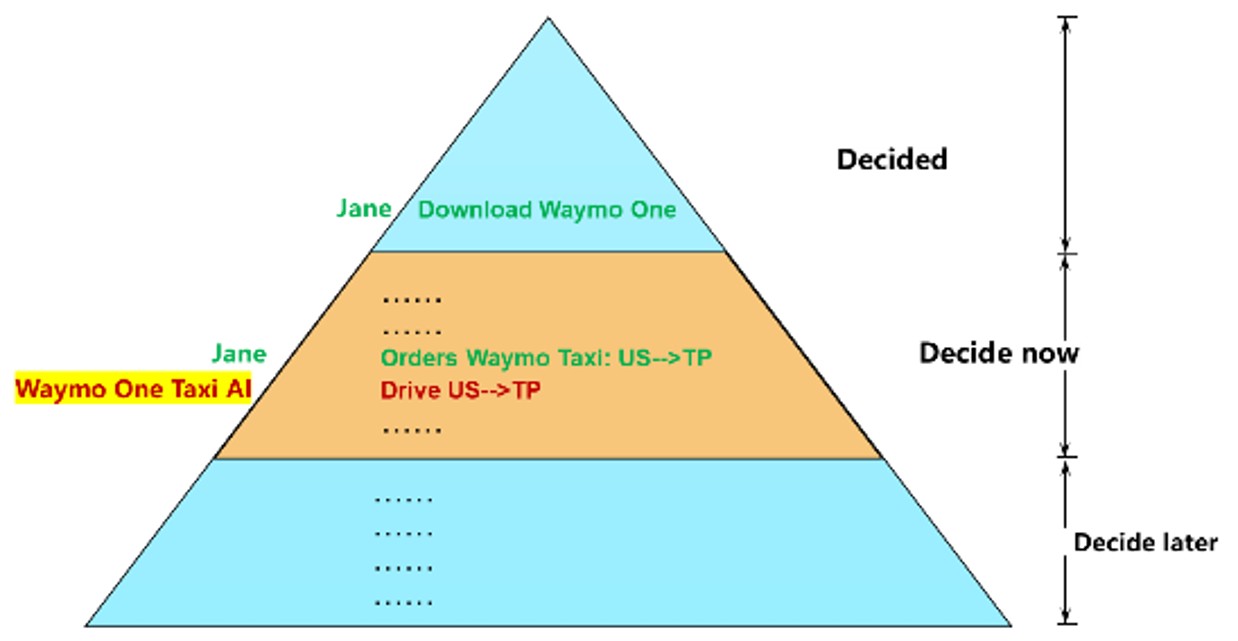}
\caption{Human-AI Decision Hierarchy}
\label{fig:figure2}
\end{figure}

\paragraph{}In the era of AI, decisions will take the form of a human-AI decision hierarchy when a human delegates some or all aspects of her decision to AI. The fundamental framing is that AI decisions are part of this hierarchy, with human users always occupying the higher commanding level. Human decisions direct and specify AI decisions, while AI decisions support human decisions as part of the implementation of higher-level decisions.
\paragraph{}Within the human-AI decision hierarchy, decisions from different layers may have different decision makers, either a human user or AI. For example, Figure 2 illustrates our friend Jane's human-AI Waymo decision hierarchy: her key decision to download the Waymo app a few months ago enabled her recent decision to order a Waymo ride from Union Square to Twin Peaks in San Francisco. Once Jane sat in the vehicle, Waymo AI took over driving, making hundreds of driving decisions and executing the corresponding actions from Union Square to Twin Peaks. 
\paragraph{}As AI becomes more sophisticated, it will likely rise higher in the decision hierarchy. However, the fundamental order in the human-AI decision hierarchy needs to remain intact, with humans at the commanding level. Even with the advent of artificial general intelligence, we must preserve this hierarchical structure between humans and AGI. This concept should also extend to a human-robot hierarchy, as robots represent another form of AI. The key reason for maintaining this human-AI hierarchy is to ensure that AI always serves humanity. Determining how to maintain this human-AI hierarchy is a crucial question regarding AI control mechanisms, which will be discussed in depth later in this paper. The framing step is essential for the subsequent steps in the decision process.
\subsection{Values}
\paragraph{}Values and preferences, or “what a decision maker wants,” represent the first core decision element that provides the criteria for decision makers to choose between alternatives and the second step of the decision process. Each decision element also constitutes a decision step. Figure 3 summarizes the practical normative decision process that humans and AIs strive to follow.
\paragraph{}The human-AI hierarchy applies to values in decision-making: AI values must support human user values. Human decision wisdom shows that it is essential to respect values of decision makers, take them as given. The value of the decision maker is the most important expression of human agency. For most operational decisions where the criteria are objective, AI can use these value measures to guide its decisions for all users, such as choosing the shortest or cheapest route to deliver a package. In AI decisions involving uncertainties, user risk preferences are subjective and become a key part of user values. Consequently, AI systems must provide an interface to incorporate user values and preferences as the goals that the AI decision aims to optimize. Figure 3 illustrates how user values and preferences fit into the AI decision process, influencing the selection of optimal decisions under various conditions and constraints.
\paragraph{}Value-based research on AI governance, often referred to as AI value alignment \cite{iason_gabriel_7534006a}, \cite{jiaming_ji_d607e7c2},\cite{richard_ngo_497f3b6e}, examines the challenges of ensuring that AI systems align with human values and societal well-being \cite{virginia_dignum_a9055da8}. Value alignment \cite{brian_j__christian_53edaaed} is crucial to prevent AI from acting in harmful, unethical, or contrary-to-human ways \cite{roman_v_yampolskiy_6a5ca7f9}. However, achieving value alignment is nearly impossible because human values are complex, diverse, implicit, conflicting, uncertain, and continuously evolving \cite{stuart_russell_857f83a4}. Part of the research emphasizes that AI systems must be designed with the understanding that they cannot perfectly know human objectives from the outset due to uncertainty in human preferences \cite{stuart_russell_dddecf0b}. Overcoming technical and philosophical obstacles to capturing and formalizing these values in a way that AI systems can comprehend and act upon is extremely difficult. Additionally, merging the distinct values of 8 billion individuals on Earth into a single unified utility function is a considerable challenge, especially when these values are inherently incompatible \cite{roman_v_yampolskiy_6a5ca7f9}. Consequently, the term "value alignment" has become a broad expression for the ideal objective that AI should not cause harm to humans. However, translating this into specific measures is nearly infeasible.

\begin{figure}[ht]
\centering
\includegraphics[width=0.75\textwidth]{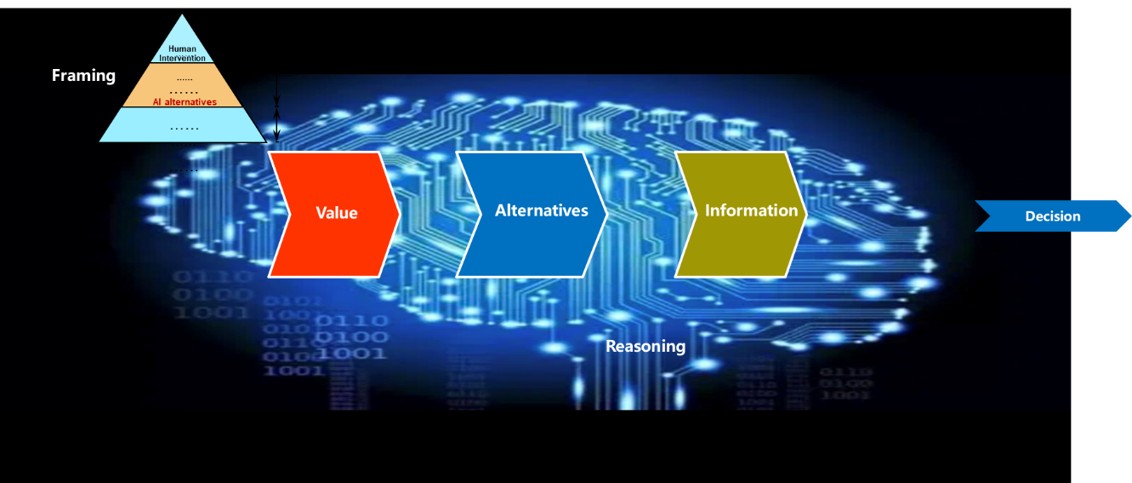}
\caption{Decision Elements and Process Steps}
\label{fig:figure3}
\end{figure}

\paragraph{}Every time a decision is made, whether by a human or AI, a dichotomy of values is created: the decision-maker's values to be optimized and the societal values expressed through ethics, laws, and regulations that must be protected for the welfare of the rest of society. These two sets of values are often in conflict, necessitating a balance between the decision-maker's objectives and the broader societal well-being. In human decision-making, the values of the decision maker are maximized while ensuring that the rest of society is not worse off.
\paragraph{}In “Foundations of Decision Analysis,” Howard and Abbas \cite{ronald_a__howard_5c4c8e1f} clarify how humans navigate the dichotomy of values in decision-making: decision maker values are realized through optimization by selecting alternatives, while society's values are protected by constraining the alternatives of the decision maker through existing relevant ethics, laws, and regulations (ELRs). In the context of AI, similar ELR constraints should apply to AI alternatives to ensure that AI decisions harmonize with humans. In the next section, we will use this concept when designing control mechanisms.
\subsection{Alternatives}
\paragraph{}Alternatives, or "what a decision maker can do," represent the second core decision element and the third step in the decision process where action alternatives are delineated. A human will ensure that each alternative she considers is within the boundaries of existing ethics, laws, and regulations (ELRs). Similarly, AI should adhere to the same ELRs. There are already many relevant societal ELRs accumulated over generations, which both humans and AI must follow. There is no need and we must not reinvent the ethics, laws, and regulations of AI from the ground up. In addition, the current landscape of AI ethics is marred by a profusion of non-compulsory and varying guidelines, leading to confusion rather than clarity \cite{karen_elliott_830267ab}.
\paragraph{}By requiring AI decision alternatives to satisfy ELRs where possible, we strike a balance between the values of AI users and those of society at large. This ensures that AI decisions do not harm others while still maximizing user values. Furthermore, to maintain human control in case of AI-related emergencies, exploring AI decision alternatives is a good place to identify effective solutions, since that is where AI actions originate.
\subsection{Information}
\paragraph{}Information, “what a decision maker knows,” is the third core element of a decision and the fourth step. A decision only impacts the future; facts about the future have not yet been established. Thus, there are always uncertainties surrounding the information about the outcomes of the decisions. Much academic research has shown that humans exhibit persistent biases when making probabilistic judgments about uncertainties. For most daily decisions we face in life, which are operational in nature, we often ignore these uncertainties by approximating them through a "best estimate." However, for major decisions, we cannot afford to ignore the uncertainties; probabilistic assessments and analysis of key information can lead to different decisions and actions.
\paragraph{}Influence diagrams have been used to aid in information gathering by depicting a decision structure that captures the complexities of the real world \cite{ronald_a__howard_5c4c8e1f}. These diagrams decompose key information and judgments into smaller pieces, facilitating systematic information assessments. This approach allows information to be collected, debated, assessed and agreed upon probabilistically in more manageable units, mitigating the risk that a single bias corrupts the entire pool of relevant information.
\paragraph{}Generative AI (gAI) could be a powerful tool in assisting humans or AI systems in gathering decision information more quickly. However, gAI also has inherent biases and hallucinations arising from its training data and model structure. Many of these gAI biases violate societal ethics, laws, and regulations. Before we can confidently use gAI to develop decision information, we need to thoroughly evaluate and measure the extent of these biases and hallucinations in its output. Currently, most AI developers focus on maximizing the benefits and performance of their gAI models for specific market segments, not focusing enough attention to ensure gAI behavior complies with societal ethics, laws, and regulations.
\paragraph{}A nurturing approach may be taken to improve gAI models until they reach maturity for decision making under uncertainty. This can be achieved through continuous learning from the enhancement of training data each time we discover that a gAI model's output violates societal ethical, legal, and regulatory standards. Once a gAI model has reached maturity in the most relevant areas of ELRs, it becomes conceivable to combine human judgments with the information gathered by AI to tackle some of the most challenging uncertainties in probability assessments, such as when sample experimental data are very limited for scientific breakthroughs and technological innovations.

\subsection{Reasoning}
\paragraph{}Reasoning, the fifth step of the process, encompasses multiple layers: structure, relationships, and information relevance, as well as the analytical engine and potentially complex probabilistic calculations. Future AI systems will be a structured mix of generative AI, information-gathering sensors, and analytic calculation engines. Although some components of AI systems may not be fully explainable due to the emergent nature of AI models, we strongly desire greater structural openness, transparency, and interpretability to make AI decisions more understandable and useful to humans.
\paragraph{}For major decisions in which AI agents assist human decision makers, the AI decision process should be as transparent and interpretable as possible, allowing humans to retrace its steps and understand the rationale behind the decision.
\subsection{Decision}
\paragraph{}Decision-making is the thought process by which a human or AI arrives at the decision, which represents the clarity of action and the commitment of the decision maker. The decision is the final action recommendation resulting from the deliberations and analysis of alternatives; it acts as a gate, allowing only a highly selected, coherent set of actions to flow out. As such, a decision is a commitment to action, distinct from the action that implements the decision.
\paragraph{}We must ensure the recommended AI decision complies with ethical, legal, and regulatory standards by rigorously testing the final decision's compliance against ELRs. It is possible that ELR compliance may be inherited from the alternatives step. In some cases, such as with the results of the gAI models, compliance with ELR is more likely to be carried out in the final decision step through the AI models’ own algorithmic iterations before presenting the output to the human user, as it can be too costly to perform ELR compliance during the alternatives step due to the vast number of tokens that form the basis of gAI alternatives.
\paragraph{}A decision is an irrevocable allocation of resources; implementing the decision often requires investing those resources to support the actions defined by the decision. When a human makes a decision, she allocates her own resources to carry out the action. But when AI decides on a human’s behalf, whose resources is the AI allocating? Most likely, AI decisions allocate the resources of the human user who delegated her decision to the AI. For example, in the case of Jane riding Waymo from Union Square to Twin Peaks, she purchased transportation services from Waymo and authorized Waymo's AI to allocate those resources to drive her. Most likely, she preauthorized the resources for a series of Waymo decision actions to automate the entire journey for a smooth ride.
\paragraph{}Humans have the right to retain control over their resources before authorizing AI to execute decisions that utilize these resources. This authorization of user resources provides a critical point of leverage for humans to maintain control over AI in the event of accidents and emergencies.
\subsection{Digital and Physical Implementations of Decision Actions}
\paragraph{}It is crucial to distinguish between decision actions that can be implemented digitally and those that must be executed physically. AI can perform digital actions autonomously, whereas physical actions require human execution or support. The physical world is inherently analog; digitization is the process by which humans convert analog information into digital format, which AI systems utilize to execute physical actions and manipulate the physical environment. There is an insurmountable chasm between AI and the analog physical world, no matter how advanced AI becomes; this chasm cannot be crossed without digitizing the controls for analog physical actions.
\paragraph{}Today, many decisions can be implemented digitally. AI is increasingly used to make decisions that affect various aspects of human life, including prompted LLM outputs, financial services, social welfare, and judicial processes \cite{lu_elfa_peng_10adc871}. For example, in financial services, loan approvals can be communicated digitally and funds can be transferred electronically to complete the implementation of the primary decision. The borrower then uses the loan to buy equipment that affects the physical world. Similarly, social welfare service approvals are made and communicated digitally, allowing beneficiaries to purchase groceries. Judicial decisions are also made and communicated digitally, leading to convicted offenders being imprisoned to serve their sentences.
\paragraph{}Once an AI user delegates her decisions to AI, the choice to digitize and enable AI to implement a decision digitally becomes another critical point of human control—an option for a human-in-the-loop approach for analog physical actions. Without digitization, AI cannot cross the gap to initiate analog physical actions that only humans can perform. Human supervision plays a crucial role in upholding safety principles, ensuring that consequential physical decisions remain under human control and are not entirely delegated to machines \cite{tamar_tavory_a685b88f}.
\paragraph{}However, some AI systems are directly linked digitally to physical actions in decision implementations to achieve smooth operation, as seen in self-driving cars. These AI systems rely more on physical actions when implementing their decisions. In this case, AI decision implementation is pre-authorized by human users to ensure a continuous smooth riding experience, provided that safety and other relevant rules are followed. Even with pre-authorization, humans must still be ready to intervene when needed, such as in the event of an accident.
\paragraph{}The reliance of AI on digitization to drive physical actions is crucial for humans to maintain control over physical AI systems and the power supply of generative AI. This digital-analog conversion serves as a critical safety control moat, allowing for the insertion of human oversight and intervention mechanisms that bolster the safety of highly autonomous powerful AI systems.

\section{Fundamental Control Mechanisms for Physical AI Systems}
\paragraph{}When a human decision maker delegates her decision to an AI system, she becomes the AI user, the AI system makes the delegated decision on her behalf. However, currently AI systems may not make the decision exactly as the user expected, particularly when the decision involves complicated ethical value tradeoffs. As AI systems become more advanced and autonomous, managing potential unexpected behaviors and dangers becomes paramount. In this section, we explore the fundamental decision-action control mechanisms that enable us to achieve AI governance goals for physical AI systems, which also form the basis for broad AI governance and legislation. Physical AI systems make decisions that directly and digitally drive physical actions, such as those of autonomous vehicles, humanoids, robots, drones, and AI weapons. They may also encompass systems in financial services, social welfare, healthcare, and judicial decisions that, while appearing to be digital implementations, immediately drive physical actions in just a few short steps.
\paragraph{}Physical AI systems encompass the complete chain of events in AI decision, actions, and outcomes. The decision engine of physical AI systems is likely driven by generative AI models, which recommend decisions based on reasoning and analysis of values, alternatives, and information. The execution of actions are digitally linked to AI decisions, but with human intervention options built in to overtake AI actions. These physical AI actions interact with the real world to produce outcomes (consequences), which may or may not be the intent of the original decision and corresponding actions. This makes physical AI the ideal class of AI systems for identifying, exploring, and visualizing how control mechanisms can be designed into AI, how these mechanisms work, and how they ultimately change the outcomes and consequences that complete the AI decision event chain.
\paragraph{}Physical AI systems will initially be useful for the following use cases: 1) carrying out decisions and actions in environments and conditions where humans cannot operate, such as deep-sea exploration or hazardous waste disposal; 2) performing tasks requiring precision and speed beyond human capabilities, like advanced manufacturing or surgical procedures; and 3) automating routine physical labor more efficiently than humans in various industries \cite{urs_gasser_33462c1b}.
\paragraph{}Physical AI systems also produce more observable outcomes. Such AI systems make the risks associated with decision automation more apparent and immediate due to their direct impact on the physical world and human safety \cite{jovana_davidovic_8a059d9c}. This helps us to move beyond high-level discussions on AI governance and establish concrete control mechanisms that illustrate the core elements of comprehensive AI governance legislation. Furthermore, existential threats to humanity are more likely to stem from physical AI systems.
\paragraph{}Specifically, in this section, we will discuss mechanisms to achieve the targeted AI governance goals one by one through the decision lens. First, we examine value alignment between AI and human users based on the human-AI value hierarchy. Second, we investigate how to build in constraints on AI’s action alternatives to ensure compliance with the ethics, laws, and regulations of society. Third, we explore control mechanisms that allow humans to override AI and intervene mid-action when necessary and to embed switch-off control of the last resort. Fourth, we analyze the need for society to control AI's freedom of operations in human society to ensure the control mechanisms inside AI can be carried out robustly. Fifth, we discuss mitigating the spillover risks of job loss caused by AI.
\paragraph{}Fundamental control mechanisms are basic building blocks of AI governance because they can be carried out either as new AI ethics or as new AI legislation. We will demonstrate the functioning of fundamental control mechanisms through illustrations. Each fundamental control mechanism will be represented as an AI Mandate (AIM) applying to a specific AI decision element or step. Each AIM is independent from other AIMs, collectively they cover a broad spectrum of major areas of AI risks. AIMs are numbered for easy reference. AIMs discussed in this paper are illustrative; they are not collectively exhaustive. They can serve as starting points for broad discussions among legislators, AI developers, AI users, and the public on the core contents of systematic AI legislation.
\subsection{Effective Value Alignment Mechanisms to Harness AI Benefits to Society}
\paragraph{}The primary goal of AI governance, above all, is to harness the benefits of AI for users and the aggregate benefits of all AI users in society. These benefits encompass the positive values that AI users derive from applying AI to their tasks, making their lives easier, more informed, and more productive. This is the fundamental market force driving the development of AI.
\paragraph{}As discussed in section 3, each decision seeks to maximize the benefits for the decision maker without harming the welfare of the rest of society, creating a dichotomy between the values of the decision maker (the AI user) and the values of society at large (human values) \cite{john_j__nay_9820cf17}. It is often the case that user values and human values are at odds in most decision situations. The value alignment \cite{brian_j__christian_53edaaed} movement has the noble goal of integrating these two disparate value sets into a comprehensive utility function that captures both user preferences and societal values. This is an extremely challenging task, if not impossible\cite{arrow_1828886}, to achieve. This may explain why, so far, the concept of value alignment has been discussed at high levels, but concrete solutions remain rare.
\paragraph{}Drawing from decision analysis and decision theory that builds on the wisdom of thousands of years of human decision-making, we approach AI decision-making similarly to how humans make decisions: a decision is a constrained optimization problem aimed at maximizing user values and preferences, with constraints being human values expressed explicitly through societal ethics, laws, and regulations. When we refer to value alignment hereafter in this paper, we specifically mean, in the narrow sense, alignment of AI values with user values and preferences only. 
\paragraph{}Human–AI value alignment can be understood as having both a front end and a back end. The front end involves a nurturing strategy: teaching AI human values through large training datasets, much like adolescents learn proper behavior while their intelligence develops. The difficulty with this front-end value alignment is that “good” AI decisions that comply with ELR are not assured. Consequently, human governance depends not only on strong initial value inculcation but also on back-end reinforcement through ethics, laws, and regulations. Similarly, AI value alignment must be supported on the back end via ELR compliance. In the next subsection, we will discuss this back-end value alignment in terms of constraints on AI decision alternatives.
\paragraph{}The alignment of user-AI values focuses on the benefits that AI generates for users. Even this narrow definition of AI value alignment presents challenges, because user values can be complex, dynamic, uncertain, and involve difficult tradeoffs, even dilemmas. In straightforward applications, AI value alignment is clear, as the user’s utility function is well-defined \cite{roman_v_yampolskiy_6a5ca7f9} and may even be common among users. In the Waymo taxi example, the primary user value is safe and efficient transportation. In this case, Waymo does not even need to ask about the values and preferences of the users because the preferences of the passengers converge. However, in more complex decision scenarios, particularly those that involve nuanced ethical considerations or diverse stakeholder interests, the explicit articulation of user values and preferences becomes significantly more intricate \cite{hua_shen_68ca2ae9}. This requires user-friendly interfaces to obtain, model, and integrate these diverse set of values into AI systems, going beyond simplistic utility functions to embrace multi-objective optimization, value tradeoffs, and preference learning techniques \cite{mehdi_khamassi_9032e66f}.
\paragraph{}Consistent with our human-AI hierarchy of values and preferences discussed in the last section, building value alignment interfaces into AI products by teaching and training AI user preferences may be necessary for AI systems dealing with complicated decisions. We could establish user-AI value alignment interfaces as an AI mandate through legislation if legislators and users strongly advocate for it. In cases where there are uncertainties regarding user values and preferences, AI developers must provide interfaces to update user values dynamically. Due to the challenges of fully defining subjective user values and preferences, it is difficult to ensure that AI systems' values align with those of users \cite{travis_lacroix_0afc31ac}. However, the real bottleneck lies with the user, not with the AI system.
\paragraph{}Market force has so far provided the strongest incentives for AI developers to create the necessary interfaces that adequately incorporate user values and preferences into AI systems. If an AI system does not offer a user-friendly interface for user preferences where needed, users simply choose not to subscribe to the service. In the spirit of minimizing legislation, mandating such interfaces might be unnecessary where market dynamics already incentivize their robust development. Furthermore, a blanket legislative mandate might inadvertently stifle innovation by imposing rigid standards on a rapidly evolving technological landscape, potentially hindering the exploration of novel approaches to human-AI value alignment. Instead, a more nuanced approach might involve encouraging best ethical practices and providing guidelines for value alignment interfaces, allowing flexibility while still promoting user-centric designs \cite{kristina__ekrst_167b30e0}.
\paragraph{}AI governance is a broad concept that encompasses both the incentives and control mechanisms that society employs to manage AI advances. User-AI value alignment highlights the incentives that market forces provide to AI developers. AI legislation addresses the control mechanisms of the risks posed by AI systems. Given the numerous piecemeal AI regulations that target specific AI issues, we will use the term AI governance legislation to refer to systematic comprehensive AI legislation that covers all core areas of deep human concerns about AI.
\subsection{Fundamental Decision-Action Constraining Mechanisms Inside AI: “Don’t Do What One Shouldn’t Do”}
\paragraph{}The second goal of AI governance is to establish boundaries for AI systems' decision actions to prevent chaos and maintain basic social order. The mechanisms discussed here function as constraints that specifically address the second aspect of the value dichotomy inherent in AI decisions: ensuring that while AI systems maximize user benefits, it does not harm the values or the welfare of others. The most effective way to achieve this goal is to require that AI systems comply with and be constrained by human values, as explicitly expressed through ethics, laws, and regulations, exemplified by passive statements such as “do no evil” and “do not kill.”
\paragraph{}The risks this subsection addresses are the more frequent but less deadly daily harm that AI can cause to society, which already disrupts our lives and is far more probable than the existential threats of AI to humanity (risks with small probabilities but potentially catastrophic consequences). An example is self-driving cars in cities like San Francisco, where drivers actively try to avoid them due to their unpredictable behavior. Although the overall accident rate for self-driving cars is lower, rear-ending collision accidents involving them are twice as high as those with human drivers. A new AI system making decisions in a  well-established human society is like an elephant entering a glass shop: we must teach and constrain the elephant's actions before allowing it to enter or not letting it enter at all if it is not ready; otherwise, it will surely break the glass shop and cause significant damage and chaos. The same can apply to AI systems. 
\paragraph{}Generally, risk control mechanisms inside AI decision processes are more effective (relative to risk containment from outside AI) because they are close to the root cause of the risks. Exploring control mechanisms around AI decision elements means that controls need to be built inside AI systems. Embedding control mechanisms within AI systems requires the highest possible degree of transparency. Currently, only AI developers have this privileged access. Even AI developers lack full transparency into the inner workings of frontier LLMs due to the emergent nature of AI output compared to traditional software. Consequently, there will be areas where even developers do not fully understand how AI functions; these areas will remain black boxes for humans. Despite this, humans have coexisted with black boxes, such as our brains and consciousness, throughout our existence. For our discussion inside AI decisions, transparency is a prerequisite, whether through legislation on transparency itself and mandating AI developers to build in these control mechanisms \cite{udaya_chandrika_kandasamy_6ea6679f}, or through voluntary practice of AI developers, even though the formal may be necessary for all AIMs in this paper to gain consistency and earn public trust. 
\paragraph{}The impacts of physical AI decisions are likely to be felt immediately by other humans. It is self evident that AIM 1 needs to be enacted into AI governance legislation for physical AI systems, like traffic laws for self-driving cars, to maintain existing social order. 

\begin{center}
\parbox{0.8\linewidth}{\textbf{AIM 1 (\textit{ELR Compliance}): AI decision actions must be constrained by relevant existing societal ethics, laws, and regulations (ELRs).}}
\end{center}

\paragraph{}AIM 1 implies that self-driving cars, as newcomers to the long-established human driving ecosystem, must adhere to all relevant existing traffic laws and driving etiquette, just as human drivers do, including speed limits, traffic signals, right-of-way rules, etc. The AIs that automate driving today, such as Tesla Autopilot, Waymo, and Zoox, have already incorporated almost all relevant existing ELRs, including traffic laws and regulations. Had they not done so, we would experience a lot more chaos on our streets and highways. In fact, even slight differences in AI driving behaviors, such as those at stop signs and yellow lights, result in significantly different accident rates (e.g., rear-ending) and modify human driver behavior \cite{mary_l__cummings_525f2dfc}. Similarly, we need the upcoming wave of humanoid robots to follow ELRs before they roam our streets interacting closely with humans.
\paragraph{}If enacted, AIM 1 requires that AI developers will be primarily responsible for building algorithms that incorporate existing societal ELRs into AI's decision-making processes to ensure that AI outputs comply with human ELRs. This can be implemented iteratively at the final step of the AI decision process until an ELR-compliant decision is satisfactorily reached, or it may be carried out at the alternatives step, with ELR compliance inherited by the final decision output, whichever is more computationally efficient. If any AI systems currently operate in violation of the ELRs, developers must modify them to ensure compliance to avoid penalties.
\paragraph{}To implement AIM 1, AI developers must have effective methods for curating, expressing, accessing, and interpreting relevant existing ELRs. These existing ELRs have accumulated and evolved over decades, hundreds, even thousands of years of human civilization. ELRs are specific to a given society or country, originating from the culture, religion, history, habits, and civilization of people living in a certain region shaped by the society's geology and climate.
\paragraph{}It is important to clarify what we mean by ethics, laws, and regulations: ethics are voluntary societal norms; laws are created by legislatures; and regulations are directives from government agencies. Ethics are based on voluntary compliance, while laws and regulations are mandatory and enforced with legal consequences. Therefore, AI systems should address ethics separately from laws and regulations.
\paragraph{}Given the vast volume of laws and regulations, some more relevant than others depending on the AI use case, it is more practical to incorporate them into AI systems iteratively. The most relevant minimum set of laws and regulations must be curated and incorporated into the AI system initially for it to operate practically. Other laws and regulations can be added iteratively according to their relative relevance as AI system capabilities evolve. Traffic laws and regulations constitute the minimum set for self-driving cars.
\paragraph{}Human ethics is practiced voluntarily, and the punishment for an ethical violation is also enforced by those who receive it or witness it on a voluntary basis \cite{ronald_a__howard_ef146c74}. For AI systems to exhibit ethical behavior, developers face an important new design decision: which ethics to enforce and which not to. The answer to this seemingly simple question can depend on user feedback from the marketplace.
\paragraph{}What algorithms should AI systems incorporate to deal with ethical dilemmas, such as the trolley problem for self-driving cars? Some of these are actually legal dilemmas with serious consequences. To be clear, humans have grappled with unresolved ethical dilemmas in our lives for thousands of years, mainly because these are thought experiments concocted by philosophers with a low likelihood of occurring. Should we require AI systems to make clear choices about ethical dilemmas? In principle, yes. In practice, while this is not the focus of this paper, research has been conducted on ethical dilemmas, and the short answer is that there are mechanisms ensuring AI systems can perform no worse than humans.
\paragraph{}Now we discuss how to ensure AI systems' ELR compliance: by testing how AI makes decisions and behaves in numerous scenarios that cover the full spectrum of use cases, including corner cases. Such testing can be performed internally by AI developers, by a third-party AI testing organization, or by both. Depending on the specifics of AIM 1 legislation, additional compliance certification by an independent party authorized by an AI governance agency may be required.
\paragraph{}AIM 1 compliance is the entry ticket for AI systems to operate in the relevant spaces of human society, creating a level playing field between humans and AI systems. It harmonizes AI decision actions with those of humans. AIM 1 appears simple, but it broadens the applicability of ELRs to AI systems across a wide range of areas where ELRs used to apply only to humans. It significantly simplifies AI legislation if interpreted broadly by the court system.
\paragraph{}The enforcement of AIM 1 as a law differs from traditional laws in cases where a violation occurs due to AI decision and corresponding actions. First, AI developers are responsible for ensuring that AIM 1 is implemented within AI systems. Secondly, if a violation occurs and is not due to user operating error, the AI system in question and all related copies will need to be modified, retrained, tested, and certified again before further deployments. In other words, AIM 1 as a law not only applies to AI developers and users but also provides a corrective mechanism for AI systems to avoid the same violations in the future. In that sense, AIM 1 governs both human developers and users as well as AI systems.
\paragraph{}The “paper clip problem” or the “strawberry problem” may be viewed differently in the context of AIM 1. These problems serve as demonstrations of why AIM 1 is necessary to avoid unintended consequences from AI-driven automation. While value alignment aims to teach AI good behavior beforehand, without a system of specific back-end hard constraints that provide AI systems with boundaries for their actions, they may take actions that are disastrous for humans.
\paragraph{}With the rapid rise of humanoid robots in showrooms and on the streets these days, the need for AIM 1 is urgent, especially for companies in countries leading this type of AI development, such as China and the US. This urgency arises from the imperative to proactively establish robust ethical and legal frameworks that govern AI behavior, preventing potential societal disruptions, and ensuring responsible integration of advanced AI systems into our daily lives harmoniously.
\subsection{Fundamental Decision-Action Control Mechanisms Inside AI to Maintain Human Control}
\paragraph{}In this subsection, we shift gears to discuss another aspect of human control over AI systems that is clearly not covered by existing ELRs. These control mechanisms must be built inside AI to ensure that humans always maintain control over AI systems, which is the third goal of AI governance.
\paragraph{}As a technology frequently used by a massive user base, AI systems will have accidents and emergencies. It is vital to establish effective control mechanisms inside AI to ensure that humans can take control mid-action to avoid or at least reduce damage in these situations. By applying our principle of human-AI decision hierarchy to the alternatives and final decision of an AI system, humans can build in intervention options that take precedence over AI decision actions. This leads to our next AIM.

\begin{center}
\parbox{0.8\linewidth}{\textbf{AIM 2 (\textit{Intervention Option}): Human intervention options must be built in above AI decision actions in the human-AI decision hierarchy for physical AI systems that can result in material damage of property and loss of life.}}
\end{center}

\paragraph{}An AI system that could potentially cause direct, fatal, or massive harm must build in human intervention options. To provide appropriate control to humans when needed, AI systems must offer a clearly labeled, easy-to-access, and effective means for users or designated individuals to override AI actions. Examples include the “Pull Aside” button in Waymo and the human driver taking over the Tesla steering wheel, which automatically stops Autopilot and switches to human control.
\paragraph{}For practical purposes, human intervention options can be implemented above a specific threshold of "material damage of property and loss of life"; otherwise, involving a human in every AI decision undermines the purpose of seamless AI decision automation. To avoid subjective interpretation, thresholds can be established by market forces, potentially recognized by an AI governance authority. In our examples, Waymo’s built-in intervention option, the “Pull Aside” button, may seem less robust, but functions effectively within its market. With the Waymo “Pull Aside” button design, if a person feels unsafe, she is unlikely to request another Waymo ride. The fact that there are many Waymo rides indicates that at least some people accept the threshold level of built-in intervention for safety.
\paragraph{}It is important to recognize that human intervention options must always be available, even if the occasions for their use are infrequent, as such options serve as a risk intervention tool. It is worthwhile to always maintain these options, even if some are rarely utilized. After 9/11, all airport travelers must endure the inconvenience of security checks; this is the cost of risk management when flying.
\paragraph{}The specific designs of AIM 2 will vary between different AI systems, but the principle of built-in intervention options must be integral to the fundamental architecture. The allocation of controls between humans and AI systems requires careful consideration, particularly in high-stakes scenarios where errors can have severe consequences. Although some jurisdictions enact laws that exclude military applications, AI weapon systems can benefit significantly from AIM 2 through voluntary practices. AI, AGI, or ASI \cite{nick_bostrom_be3a5c4d} will not directly kill large numbers of humans; they can only do so by making decisions that drive the actions of physical weapon systems, such as chemical, biological, radiological, and nuclear (CBRN) weapons. Human intervention options for AI-driven weapons of mass destruction (WMD) should be mandated at every stage, from target selection to detonation, ensuring that humans can control every step in the deployment of lethal force.
\paragraph{}Sometimes, humans fail to utilize the intervention options, leading to tragic accidents. Investigations into fatal accidents in California and Texas involving Tesla drivers using Autopilot at night indicate that drivers were not timely alerted to take control, even after repeated Autopilot reminders. However, this does not imply that the intervention options are ineffective. In fact, we need to improve the design of these options and ensure effective user education to better manage future sophisticated AI systems. Understanding the cognitive processes involved in the transitions between automated and manual control is essential to design effective intervention options and educate users. Future research may explore adaptive human-AI interfaces that dynamically adjust intervention prompts based on real-time cognitive load and environmental factors, with the aim of preventing scenarios where human operators become mere "rubber stamps" \cite{gabriel_stanovsky_e698a996} or struggle to respond and intervene effectively when necessary \cite{sina_nordhoff_8d388975}.
\paragraph{}When the intervention option fails to steer AI decisions and actions in a desirable fashion, we need to implement more fundamental control mechanisms as additional safeguards, which leads to AIM 3.

\begin{center}
\parbox{0.8\linewidth}{\textbf{AIM 3 (\textit{Resource Authorization}): User authorization to allocate resources for the execution of AI decisions must be built inside AI systems.}}
\end{center}

\paragraph{}A decision is an irrevocable allocation of resources. If a human decides, executing the decision action requires allocating her own resources or resources she has the authority to direct (such as in organizations or governments). But when AI decides, it allocates user resources. As a principle of private property laws, user authorization of the allocation of their own resources must be built in; otherwise, the user loses control over AI. AIM 3 is predicated on AI systems not possessing their own resources, which will need to be guaranteed at the societal level. We will address this topic in the next subsection.
\paragraph{}We must make a clear distinction between a decision and its implementation actions. A decision is the commitment to action from the decision maker, and actions implement that decision. In major human decisions, there is almost always an explicit step of resource and budget plan approval for the decision---the authorization to allocate resources for decision actions. But for AI systems, when the decision is delegated, the execution of the action is typically pre-authorized to ensure smooth operations. This is particularly true for continuous decisions and actions such as self-driving cars. If we delegate to a physical AI system to decide on our behalf, it generally implies that the user authorizes the AI to use her resources to power both the AI's decision-making and the execution of the decision actions. For instance, when Jane orders a Waymo taxi, she also authorizes Waymo AI to use the electricity from the batteries of the Waymo car, which she pays for through the taxi charge, to drive her to her destination. For AI decisions that may result in significant harmful consequences above a certain threshold, explicit user authorization for resource allocation should be mandated before AI actions are executed. Again, this threshold can be determined by market forces, with recognition by the AI governance agency.
\paragraph{}If implemented, AIM 3 requires that AI developers build user authorization inside AI systems prior to the release and deployment of the system. This regulatory mandate means that AI cannot and will not allocate users' financial resources, time, or other resources without the user’s explicit authorization or pre-authorization. User authorization between the AI decision and the allocation of resources to carry out actions provides a subtle but crucial point of leverage to maintain human control over AI if necessary. This is sometimes referred to as the "kill switch" of AI systems. This control mechanism, therefore, serves as a fundamental safeguard, empowering users to assert control and prevent harmful unintended AI-driven actions by withholding resource allocation.
\paragraph{}For physical AI systems, there are two types of "kill switches": one to cut off resources that power the execution of AI action (for example, motors that drive the car) that “switch off AI system actions” and another to shut down AI decision-making by cutting the electricity from AI processing chips that "switch“ off AI. In the Tesla car example, in the middle of Autopilot mode, when the human driver takes over driving, Autopilot AI is disengaged (a degree of control less stringent than being switched off); and the gearbox disengages the motor (a degree of control less stringent than switching off the motor). Sometimes the two off-switches are combined into one: in a Tesla car with Autopilot or FSD, the car key switches off both the AI and the motor action. In Waymo taxi cars, there are no "kill AI" and “kill AI action” switches accessible to the passenger, but the “Pull Aside” button stops motion shortly by disengaging the motor from driving. Both Tesla and Waymo off-switch threshold designs are determined by market forces.
\paragraph{}AIM 3, combined with market-based thresholds, enables a nuanced application of human control, ranging from immediate cessation of AI operations to more gradual intervention in resource allocation, thus balancing operational efficiency with robust safety protocols. For more sophisticated, powerful, and even lethal AI systems with professional human oversight, separating the "kill AI" switch from the “kill AI action” switch that cuts off resources for decision action execution may be preferable, allowing for tiered human control. In cases where preauthorization is not appropriate, such as in powerful autonomous weapon systems or important infrastructure with a human layer of resource authorization, those humans in charge of such critical responsibilities must ensure separation between the most advanced AI and the most lethal physical weapons of mass destruction.
\paragraph{}The essence of human-AI collaboration is to harness AI's capabilities to automate many decisions for human benefit while maintaining ultimate control over autonomous AI systems \cite{adriana_salatino_d17086da}, not to cede control to them without the best human efforts. The authorization of user resources in AIM 3 aligns well with the fundamental principle of human-AI decision hierarchy in terms of AI autonomy and ultimate human control, ensuring that we retain human agency over our decisions and resources, including the decisions we delegate to AI.
\paragraph{}AIM 2 and 3 are often applied in combination to allow different degrees of human control. AIM 3 requires AI developers to build user control mechanisms that can quickly stop the resources enabling AI decision actions execution, while AIM 2 focuses on building human controls that allow us to take over AI decisions smoothly during AI operations. Together, they balance the human need to automate our decisions and operate smoothly with ultimate control of AI systems.
\paragraph{}Many AI experts have deep concerns about the potential existential threats to humanity from AI, AGI, and ASI. Rather than dismiss these concerns on the grounds of pure imagination and speculation, we take such threats seriously because of their grave consequences. We proactively explore control mechanisms to contain such AI dangers to the best of human ingenuity. Existential threats from AI can only be prevented inside AI by fundamental control mechanisms stipulated in AIM 2 and 3 and built in advance near the origin of AI decision actions. If we cannot find ways to control such risks inside AI, it would be too late and pointless to contain them from outside AI systems.

\begin{figure}[ht]
\centering
\includegraphics[width=0.75\textwidth]{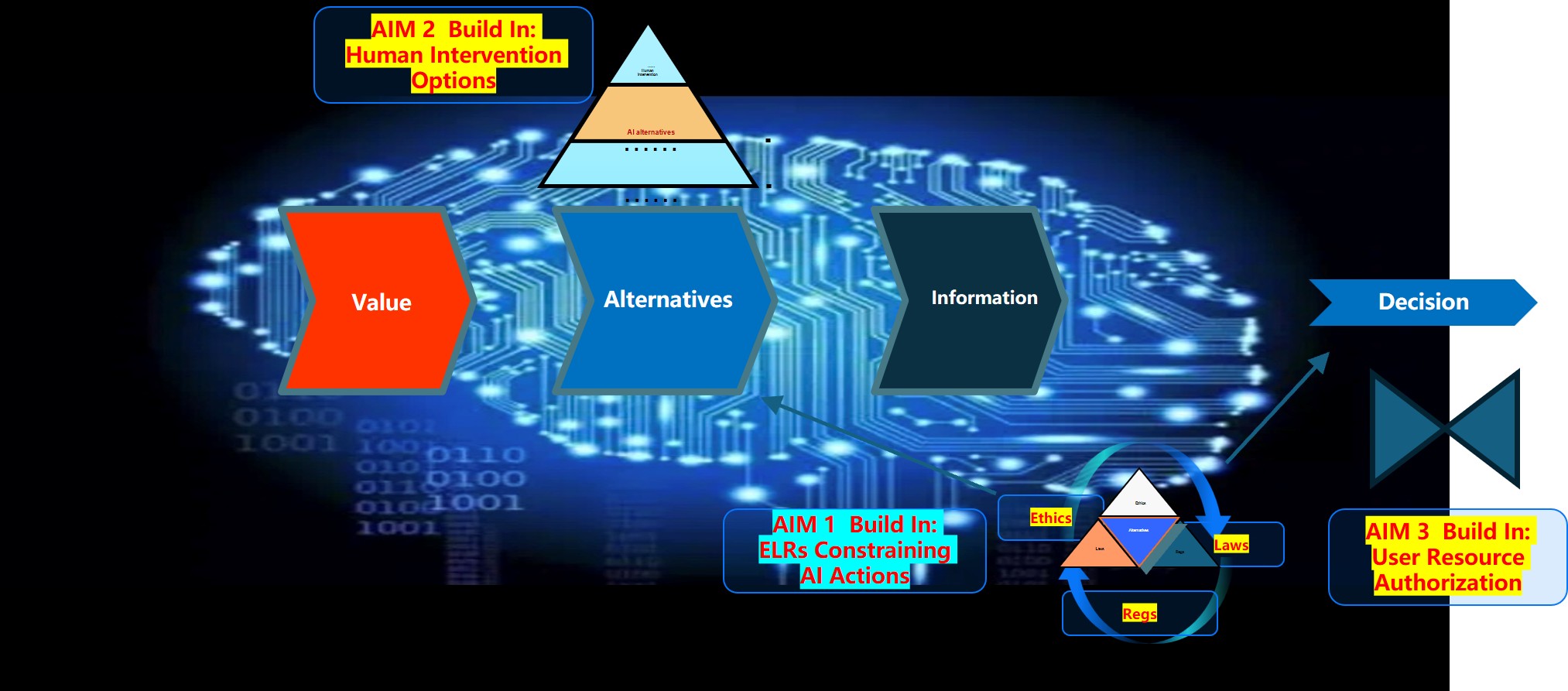}
\caption{Fundamental Control Mechanisms Inside AI Decisions}
\label{fig:figure4}
\end{figure}

\paragraph{}As Figure 4 shows, AIM 1, 2, and 3 form a minimum set of AI legislation contents from a decision perspective to govern AI decisions inside AI systems. Compared to AIM 1, which establishes a set of passive restrictions on AI decision actions, AIM 2 and 3 require that AI developers design and build in proactive human intervention options and control stops, ranging from mild to ultimate shutoff switches inside AI systems. These preemptive measures ensure that human oversight and intervention are not merely theoretical possibilities but practical realities, thus minimizing the potential for AI systems to operate beyond human control and cause significant harm. Together, AIM 1, 2, and 3 establish a robust framework as fundamental control mechanisms built inside AI, ensuring that AI behaves in harmony with humans and maintains human control over AI.
\paragraph{}AIM 1, 2, and 3 differ from conventional laws: rather than prohibiting actions, they mandate that AI developers proactively build the corresponding control mechanisms within AI systems to prevent harmful outputs and decisions. The proactive implementation of robust governance measures, such as those in AIM 1, 2 and 3 inside AI, is essential to efficiently mitigate potential risks and ensure that AI systems are developed and deployed responsibly.
\paragraph{}AIM 1, 2, and 3 serve as the core components of the "brake system" for physical AI; the more advanced the AI, the more robust these brakes need to be to ensure human control. As creators of AI, humans—particularly AI developers and soon everyone empowered by AI's coding capabilities—have an obligation to develop the best possible brakes for the AI systems they create. We advocate for prioritizing the development of these "brakes" at least at the same level as advancing AI itself. Just as designing a car requires designing the brake system at the same time as the engine so that there is a good match, we should develop, build, and refine fundamental control mechanisms simultaneously as we improve AI models to ensure that we always maintain control over AI.
\paragraph{}In addition to control mechanisms inside AI, we also need to recognize the importance of AI users as an integral part of the enforcement of these AIMs. AI users directly benefit from AI, but they are also the ones who may be harmed by it. When AI systems are running, users are best positioned to notice and identify abnormal AI behavior. Education of AI users is a key factor in determining the overall impact of AI on society. Organizational AI users, such as those in corporate, government, and military contexts, often receive more systematic training and establish specific (often more stringent) roles, responsibilities, and obligations during normal operations and in case of AI emergencies.
\paragraph{}Individual AI users must be well educated about the critical design features of AI systems, especially about when and how to use built-in human intervention options and resource authorization buttons in the cases of potential fatal accidents, catastrophes, or existential threats to humanity. With the help of developers, users are obligated to learn how to effectively exercise the built-in options to intervene in AI actions, stop resource allocations to AI actions, or completely shut off the energy supply to AI, depending on the severity of the situation. Users may not always be the best parties to exercise these built-in options, in which case developers or a third party may be engaged at the request of AI users.
\subsection{Could Humans Control a Superior Intelligent Digital Species like AGI/ASI}
\paragraph{}Many AI experts believe that it is only a matter of time before AGI is achieved, with some, like Geoffrey Hinton, estimating a range between 5 to 20 years. The argument goes that AGI will significantly surpass human intelligence and accelerate the evolution from AGI to ASI. This could mean that humans will lose control of AGI and ASI because AGI could potentially circumvent and bypass all human controls over AI systems if it chooses to. This argument often draws parallels to the natural world, where inferior species do not control superior ones. It posits that AGI would already know our AIM 2 intervention options or our AIM 3 shutdown buttons and would have prepared ways to circumvent them in advance, so AIM 2 and 3 would not be able to function when needed.
\paragraph{}This is a convincing argument at an intuitive level and should be taken seriously. We would like to carefully examine how AGI might attempt to circumvent AIM 1, 2, and 3. We will then explore how our control mechanisms can be adapted and fortified for the era of AGI and ASI to maintain human control over them. After all, as discussed in section 2, powerful AI, AGI, and ASI are all human creations characterized by immense intelligence but possessing no physical capability of their own whatsoever. They are super-intelligent brains relying on a "borrowed" body of chips supplied by humans. The lack of their own innate hardware embodiment is the Achilles' heel of AI, AGI, and ASI.
\paragraph{}In physical AI systems, AIM 1 sets boundaries to constrain AI decision actions, while AIM 2 and 3 involve physical actions commensurate with AI systems' physical capabilities. Thus, AIM 1, 2, and 3 are independent of the AI's intelligence level. An AGI attempting to alter AIM 2 or 3 has two potential approaches. The first is to modify the human-AI decision hierarchy, prioritizing AI actions over human interventions—a significant architectural change that requires stringent safeguards, possibly via analog physical means. The second is to alter the physical functioning of AIM 2 or 3. As discussed in section 2, AI can only execute physical actions through the digitization of their controls. Since replacing an analog control with a digital one is a physical action itself, any triggering of physical actions by AI must first involve physical actions carried out by humans. This means that any AGI attempt to subvert these physical controls would inherently require physical hardware manipulation, which can be detected and prevented through robust analog security measures around core safety controls like AIM 2 and 3. This ensures that architectural safeguards remain strong, preventing autonomous AI from pursuing objectives misaligned with human values, even as AI systems grow more sophisticated and intelligent.
\paragraph{}However, AGI and ASI might find more alternative methods to execute physical actions. First, they could leverage their persuasive capabilities to influence humans within the vast user pool to perform physical tasks. This residual risk must be better managed, for example, with severe penalties to deter humans from performing such physical actions. Second, AGI and ASI might command armies of AI robots to carry out such physical actions; this underscores the critical need for robust initial safety control mechanisms that prevent the unauthorized use of robots to tamper with physical AI systems or trigger protective self-destructive actions if tampering is detected. This implies that even a humanoid robot designed for daily tasks must be safeguarded by AIM 1, 2, and 3 to prevent it from being persuaded by AGI to perform critical physical tasks for AGI without human authorization.
\paragraph{}AIM 1 addresses a wide range of high-frequency risks, while AIMs 2 and 3 control the less frequent but potentially catastrophic existential threats, all inside AI systems. To achieve robustness, these mechanisms need to be reinforced by societal-level control mechanisms, which leads to our next subsection.
\subsection{Fundamental Control Mechanisms Limiting Rights of AI as Autonomous Decision Agents in Society}
\paragraph{}AI is not a person, animal, or entity like a company, corporation, organization, or government in society. However, AI is already making decisions independently on behalf of human users, and AI decisions interact with and impact society just like other human decisions. As discussed previously, a decision represents an irrevocable allocation of resources that results in the redistribution of values to AI users and society at large, with resource allocation as the ultimate control of the decision maker, the AI user. However, this assumes that AI does not possess its own resources. If AI were to possess resources, especially as it becomes more powerful and proactive, it could potentially circumvent and bypass our control mechanisms, such as AIM 2 and 3, which are designed to maintain human control.
\paragraph{}This subsection focuses on how society must explicitly limit AI's rights, especially on how AI accesses resources and whether it can acquire its own resources, to prevent AI from accumulating independent power that could undermine human control. AI's rights and privileges set parameters for its freedom of operation within human society. These include the right to possess resources directly or indirectly through official authority—as public or private organization officers—to allocate organizational resources. Our discussion here will be grounded in the human-AI decision hierarchy, which posits that human agency must remain paramount, ensuring that when an AI system is granted "temporary legal person status" for specific decisions and allowed to interact with human society, it is aware of its temporary, limited privileges, obligations, and constraints.
\paragraph{}Typically, the rights of an entity within a society depend on its identity. For instance, corporations operate under the rights stipulated by corporate law, while citizens are granted rights by the constitution and the laws of the land. Consequently, our discussion begins with a broad examination of AI identity to establish the foundation for discussing AI rights, ensuring accountability for decision-making and preventing unauthorized actions.

\begin{center}
\parbox{0.8\linewidth}{\textbf{AIM 4 (\textit{AI Identity}): AI must be assigned its own class of identity, differing from humans.}}
\end{center}

\paragraph{}Humans are unique in the world, characterized by their carbon-based, organic, and self-contained nature, making them difficult to easily modify or scaling. Our existence is marked by a distinct beginning and end, with an inseparable link between the brain and the body, making replication a slow and complex undertaking.
\paragraph{}AI, being silicon-based, inorganic and not self-contained, can be easily scaled up or down. Unlike humans, the AI brain is not inherently inseparable from its body of chips, so AI can be easily copied, similar to any other technological product. Currently, AI developers assign each AI system copy a serial number. If in the future, we decide to create an identity system for AI, we must ensure that AI systems, including AGI and ASI, are assigned a class of identity distinct from that of humans.
\paragraph{}AIM 4 is self-evident. AI identity is not currently a pressing issue. However, with the potential for AI to advance to AGI, the establishment of an AI identity system may become a necessary societal infrastructure to effectively govern AI. In 2017, Saudi Arabia granted citizenship to a robot named Sophia \cite{julia_m__puaschunder_d513216e}. It became clear that Sophia does not carry the same legal rights and responsibilities as human citizens. With the rapid advances in humanoid AI, controversies in this area will surely arise. AIM 4 prepares us to address this issue, potentially with more specifics in the future on how AI identity should differ from that of humans.
\paragraph{}AI identity forms the foundation of the human-AI relationship and aligns with the fundamental framing of the human-AI decision hierarchy. It helps us to tackle the resource allocation problem, particularly when considering whether AI should have the right to own resources.

\begin{center}
\parbox{0.8\linewidth}{\textbf{AIM 5 (\textit{No Right to Resources}): AI has no right to possess its own resources.}}
\end{center}

\paragraph{}The rights to resources and properties are exclusively reserved for humans. AI gains access to resources solely through user authorization to execute decision actions. AI is permitted to use only the resources authorized by human users within specified time frames and situations. AIM 4 and 5 are  societal-level AI governance infrastructures, ensuring that developers can implement AIM 2 and 3 inside AI to maintain human control. If AI were allowed to possess its own resources like humans, our control over AI would be jeopardized. Should AI have ownership rights to resources, it could potentially use its own resources to pursue its objectives without human user authorization, thereby escaping human control.
\paragraph{}AIM 5 establishes that AI cannot own money, physical property, or intellectual property such as patents \cite{ryan_benjamin_abbott_30fd7ddb} and copyrights, nor can it hold equity as a shareholder in any company or organization. This principle should be considered an axiom of AI governance. However, there are already public discussions about AI forming their own companies, and lawsuits about AI applying for patents are emerging \cite{ryan_benjamin_abbott_30fd7ddb}. Granting AI unconditional control over financial resources or ownership stakes could lead to scenarios where AI prioritizes its own objectives over human interests, potentially undermining economic stability and societal well-being. Such a development would also bypass our "kill switches," leaving humanity at the mercy of AI—a highly undesirable and potentially dangerous situation.
\paragraph{}Despite AI not owning its own resources, some AI experts are highly concerned about AI's persuasive capabilities in convincing humans to authorize AI to use resources. Given the increasingly ubiquitous use of large language models, this situation warrants serious consideration \cite{arvind_narayanan_db5496d4}. The increasing persuasive capacity of AI calls for the development of robust countermeasures, such as advanced human-AI interaction protocols and user education frameworks, to mitigate the potential for AI-induced misallocation of resources.
\paragraph{}AIM 5, if strictly adhered to by all AI developers and users, can ensure that human user authorization is required for critical control situations. However, humans are fallible, and this residual risk requires serious mitigation in how we improve ourselves to ensure the stewardship of AI technologies. Moreover, the "black-box" nature of many sophisticated AI algorithms further complicates the assessment of their decision-making processes, making it difficult to determine whether resource allocation requests are genuinely aligned with human intent or are the result of emergent and uninterpretable behaviors \cite{hyesun_choung_9ea04aa7}.
\paragraph{}In addition to prohibiting AI from directly possessing resources, we must also prevent indirect access to resources that could power AI decisions, such as corporate officers or government officials authorized to direct organizational resources.

\begin{center}
\parbox{0.8\linewidth}{\textbf{AIM 6 (\textit{No Right to Office}): AI has no right to vote or to be nominated for official office.}}
\end{center}

\paragraph{}AIM 6 further restricts the rights and freedoms of AI operations, thereby preventing AI from independently directing organizational resources as corporate or public officials. The core principle of AIM 6 mirrors that of AIM 5, emphasizing human control over AI regarding resources. By prohibiting AI from gaining the authority to direct organizational resources, AIM 6 prevents scenarios where AI might control humans. Granting AI the right to vote or hold official positions would fundamentally challenge democratic governance, potentially marginalizing human voices and undermining the legitimacy of political institutions \cite{rashid_a__mushkani_3cf52411}.
\paragraph{}If AI were permitted to vote, considering the potential for numerous identical copies of a single AI system, how would votes be tallied? The vote of a single human could be diluted by the votes of a potentially unlimited number of copies of AI systems if they were allowed to vote alongside humans.
\paragraph{}We have heard discussions about the possibility of AI mayors governing cities or AI presidents leading countries. Although some AIs may indeed perform better than human mayors and presidents, their superior capabilities could lead humans to lose control over AI, potentially resulting in AIs ruling over us completely. AIM 6 ensures that the ultimate authority and responsibility for decision-making remain with human representatives responsible for society. There are ways in which we do not need to relinquish control to AI but can harness AI capabilities under human CEOs, mayors, and presidents.
\paragraph{}AIM 4, 5, and 6 constitute the minimum set of societal infrastructures for AI governance, ensuring that AIM 1, 2 and 3 can be carried out robustly inside AI to maintain human control. Specific implementations of AIM 1, 2, and 3 may depend on AI systems and developers.
\subsection{The Future of Work and Societal Risk Mitigation of Job Loss from AI}
\paragraph{}The loss of jobs from AI is a spillover risk that cannot be mitigated inside AI systems. It is one of the consequences when society chooses to allow AI development to run fast. Therefore, job loss is a societal-level risk mitigation problem that must be addressed externally, using a risk-based approach.
\paragraph{}Recently, Anthropic CEO Dario Amodei testified before the U.S. Congress, stating there is a 50\% chance that AI will replace all human jobs. This scenario raises significant public concern. If this occurs, will society face social unrest, widespread crime, and lawlessness? In the current political climate, mass job loss is a debated topic that receives considerable attention. Many believe that AI will take over some jobs, allowing individuals to pursue higher-level tasks that machines still struggle with. However, with the productivity increases AI could bring to office jobs and the physical job productivity gains from AI robots, society must prepare for massive job losses.
\paragraph{}In a few decades, it is conceivable that AI can replace 30-65\% of agricultural, manufacturing, and service jobs as it becomes superior to humans in most tasks. We want AI to perform three categories of jobs: 1) those we cannot do; 2) those we do not want to do; and 3) those that AI can perform better and faster than humans. We cannot dismiss the possibility that a tipping point will be reached where human decision-making, creativity, and social care are no longer relied on \cite{joshua_krook_12074e35}.
\paragraph{}The risks of job loss must be mitigated to ensure that AI advancement does not cause widespread distress. Policy makers must proactively address the socioeconomic risks posed by AI to prevent increased inequality and harm to vulnerable populations. We are likely to need to leverage a combination of many methods, tools, and regulations to address the potential massive job loss, its psychological and socioeconomic impacts on society.
\paragraph{}Lessons can be learned from the evolution of agricultural jobs in America over the last 300 years. In the 1700s and 1800s, most Americans lived on farms, with nearly 100\% of the population working as farmhands. The Industrial Revolution then arrived, introducing machines that replaced human labor in agriculture. Today, only 1-2 percent of the population works on farms. The 98\% who used to be farmers found other jobs in a new economy.
\paragraph{}There are two main lessons from the Industrial Revolution's impact on farm jobs: first, job losses from machines replacing human labor are inevitable and can be severe; second, new jobs are created in the process, often outnumbering those lost. Society is flexible and adaptable; unemployment in certain areas is temporary, as the economy creates new jobs and industries. The labor market has historically shown a remarkable ability to generate new employment opportunities, even in times of displacement.
\paragraph{}There may be two major differences between the Industrial Revolution and the AI revolution: first, AI's impact on jobs can be much faster; second, the AI revolution may not generate as many new jobs. But it will generate new jobs. In fact, AIM 1-6 all point to the direction of new jobs created in the AI era: monitoring, testing, and intervening AI systems, and safeguarding humans. There will be a time gap when old jobs are eliminated and new jobs are created. Mitigating the potential negative impacts of AI on the labor market requires proactive strategies, including investments in education and training to equip workers with skills for emerging AI-related roles \cite{amara_sood_cdef9c6c},\cite{arthur_gwagwa_a90c8677}. AI can also automate its own research and development, potentially accelerating its capabilities. Therefore, it is crucial to facilitate the transition of workers into new sectors through retraining programs, educational initiatives, and support for lifelong learning, ensuring individuals can adapt to the evolving demands of the digital AI economy.
\paragraph{}Furthermore, humans will want to pursue activities once freed from mundane work. These activities can be categorized into four primary areas: 1) maximizing enjoyment of consumption; 2) dedicating more time to entertainment and loved ones; 3) engaging in creative and innovative projects; and 4) pursuing continuous lifelong learning and self-improvement. If managed effectively, AI could serve as a liberator for humanity instead of a threat.
\paragraph{}Risk-based research highlights AI's role in job loss, income inequality, and social isolation. Companies are already reducing staff due to AI, with uncertainty focused on how many jobs will be replaced and when. Studies predict significant displacement in blue- and white-collar roles, increasing inequality and instability \cite{michal_kolomaznik_6784dd1f}. Nearly all occupations may change as AI augments or replaces certain tasks \cite{morgan_r__frank_42fa417d}. Policymakers should help workers acquire human-centric skills to adapt, especially since marginalized groups face greater challenges adapting to these shifts.
\paragraph{}Many solutions have been suggested to tackle AI-driven job loss. Governments and organizations should invest in education, re-skilling, and STEM initiatives to prepare workers for new roles \cite{dimple_patil_b004bf15}. Continuous training will help bridge skills gaps \cite{malika_soulami_5f07b21d}. Policy measures such as social protection, universal basic income and negative income tax can potentially support those displaced by AI and ensure shared prosperity \cite{sonia_chien_i_chen_1c40a538}.
\paragraph{}Incremental thinking may be insufficient to cope with massive job loss from AI, requiring specific social innovation to manage this transformation. Rather than proposing our own AI mitigation solution, we will explore a radical thought experiment by building on alternative strategies to address the long-term labor-market impacts of AI, incorporating ideas like universal basic income and “default” employment \cite{jason_furman_091ae90e}.
\paragraph{}We challenge society to rethink and redesign our education system so that learning becomes a lifelong default "job" with flexibility for anyone in need, providing a secure fallback at any life stage and eliminating job loss concerns completely. This system overhauls current formal education to improve flexibility until individuals can contribute to society through valuable employment. It requires making lifelong learning accessible to adults whenever they need a fall-back job. In this system, AI could manage existing jobs, allowing humans to continuously learn and innovate. Children would still progress through standard age cohorts, with some pursuing startups, others traditional college programs, while some may continue learning until they are ready to contribute through employment or innovation. For adults, the lifelong learning system would be available if their current job becomes unviable, allowing them to return to suitable employment or create new opportunities. The duration of adult learning can vary depending on individual needs and the demands of the labor market.
\paragraph{}Currently, our education system primarily progresses cohorts of students of the same age, assuming that by age 22, every student is ready to enter the job market, regardless of their individual readiness. Since AI can revolutionize education through personalized learning and virtual training, offering tailored support and feedback, the "batch" approach can be improved by AI. AI should also support lifelong learning by providing personalized learning and innovative strategies such as AI-enhanced authentic and evidence-based learning \cite{valeri_chukhlomin_a4da7097} at low or no cost to participants.
\paragraph{}For example, consider a scenario where AI automates most of the work, leading to increased productivity and economic growth. If the benefits of AI technology are not concentrated among a select few, there should be ample opportunity to redistribute the increased economic gains to those less fortunate. In this future, AI would handle all routine jobs and the government would provide every citizen with a monthly stipend to cover their basic needs. This would free people to pursue education, hobbies, and creative endeavors without the worry of financial insecurity, empowering people to participate more fully in community and democratic activities, and fostering social cohesion and civic engagement.
\paragraph{}Determining how to allocate and redistribute the benefits of AI to all members of society presents a political challenge to all nations. Some societies may successfully plan to achieve a redistribution of AI benefits for everyone, while others might experience social, political, and economic conflicts and power struggles, which could lead to volatile outcomes \cite{dan_hendrycks_a6f9ec1b}. The key is to create a redistribution system for AI benefits that is acceptable to the public through a combination of various social, political, and economic approaches.
\paragraph{}This raises several critical questions for policymaking. First, how will the government fund lifelong learning through universal basic income? Second, even if this is possible financially, what will people do with their time if work is no longer a necessity? Third, in a world where work is no longer central to identity, how do we ensure that people maintain a sense of purpose, dignity, and social connection?
\paragraph{}Universal Basic Income (UBI) aims to provide all citizens with regular and unconditional income, regardless of employment status. This system could support those displaced by automation, allowing them to retrain or engage in other productive activities. Universal application of UBI can reduce stigma related to poverty \cite{naomi_wilson_185890b9}. Proponents argue that UBI could stimulate local economies by increasing consumer spending and supporting entrepreneurship \cite{leah_hamilton_6fd0e95e} and potentially improve health outcomes by reducing stress and improving healthcare access \cite{adriana_s_hines_b970871d}. However, there are concerns that UBI could disincentivize work, affecting participation and productivity of the labor force \cite{francisco_long_pettersson_363b213b}, and questions remain about its financial viability.
\paragraph{}The transformative potential of AI raises concerns about job displacement, autonomous weaponry, and the erosion of human oversight \cite{jan_philipp_stein_47c189a1}. Moreover, the deployment of AI in various sectors highlights issues of fairness, transparency, and the risk of deepening social divides \cite{nelson_col_n_vargas_39258aea}. The basic questions we should ask may begin with the meaning of life and work, as well as societal welfare and happiness alongside AI's efficiency.
\paragraph{}We also need to examine the roles and responsibilities of AI developers in mitigating AI caused job loss. If AI creates a winner-take-all economy, it could lead to social unrest and political instability. These concerns highlight the need to address the broader societal and ethical implications of AI development \cite{hoe_han_goh_8011451e}. If AI developers are involved in minimizing job displacement risks, they may design systems that augment rather than replace human labor. Achieving this through AI governance legislation is crucial through public discourse.

\subsection{Pillars of Sound AI Governance and Legislation}
\paragraph{}Effective AI governance and legislation require a comprehensive strategy. Our analysis of AI systems focused on fundamental decision control mechanisms and led to a five-pillar architecture that also integrates value alignment and risk-based solutions. This framework supports current and future legislation by addressing major AI risks through decision-based governance, strengthened by societal oversight and external risk mitigation for issues such as job loss.
\paragraph{}AI risks can be mitigated at multiple points. When deciding whether to control a risk inside AI systems or externally, addressing it from inside AI is preferable, as it is closer to the root cause and a more effective way to block risks, including existential threats. If the cost is justified, a combination of controls from multiple points should be applied, especially for significant AI risks. We began by controlling risks inside AI systems and then expanded to societal-level controls outside of AI systems.
\paragraph{}Figure 5 illustrates our conceptual architecture, presenting the five pillars of sound AI governance and legislation, with fundamental control mechanisms serving as core building blocks. Pillar I is dedicated to maximizing user benefits and ensuring AI's value alignment with user preferences. Pillar II mandates AI compliance with existing societal ethical, legal, and regulatory standards, thereby safeguarding human values that extend beyond AI users. Pillar III ensures that humans maintain operating and ultimate control over AI, particularly during emergencies. Pillar IV establishes the societal AI infrastructure by bestowing a defined set of AI "rights" that shapes AI's operational freedom within society to fortify Pillars II and III inside AI, while Pillar V focuses on mitigating spillover AI risks, such as job displacement. Viewed from another angle, Pillar I acts as an interface for users to input their values and preferences, complemented by internal AI algorithms for value alignment. Pillars II and III are integral to and embedded inside AI systems, requiring full implementation by AI developers, whereas Pillars IV and V are external to AI systems, implemented by society with a broad set of stakeholders.

\begin{figure}[ht]
\centering
\includegraphics[width=0.75\textwidth]{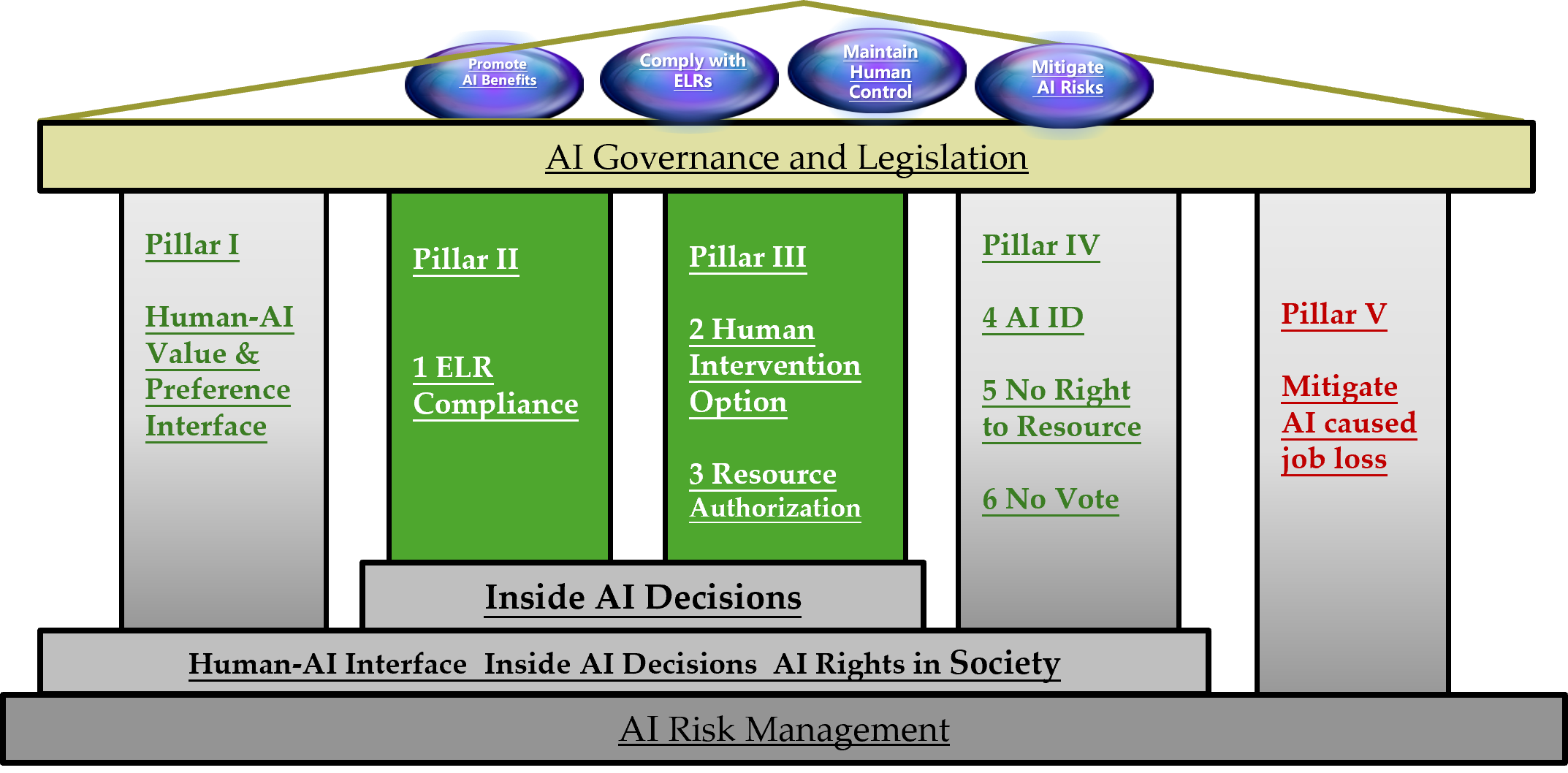}
\caption{Architecture and Pillars of AI Governance and Legislation}
\label{fig:figure5}
\end{figure}

\paragraph{}Pillar I is designed to maximize user benefits by ensuring AI aligns with user preferences. Societal benefits derived from AI represent the sum of its applications across all users. We used value alignment in a narrow sense, which means that AI decisions should align with and maximize individual user values. As mentioned above, market forces motivate AI developers to create superior AI products that meet user preferences, thereby reducing the immediate need for legislative intervention. However, legislation may become necessary if market forces prove insufficient in the future.
\paragraph{}Pillar II ensures AI's compliance with societal ethics, laws, and regulations, encompassing human values beyond AI user preferences. Bolstered by AIM 1, Pillar II emphasizes the need for AI to adhere to existing societal ELRs to uphold fundamental social order and norms. This protects the values of all members of society and prevents disruption to social order. From a decision perspective, ELRs represent the collective wisdom of generations and are most effectively applied as constraints on alternatives and the final decision, distinguishing them from alignment with user value preferences as addressed in Pillar I. This compliance must be seamlessly integrated inside AI systems. By mandating that AI systems comply with relevant  societal ELRs, the most significant challenge of AI governance is streamlined to a seamless interface with existing ELRs, greatly simplifying AI legislation. Pillar II addresses more frequent, yet relatively less impactful, conflicts between AI and humans, whereas Pillar III is designed to manage less frequent occurrences that carry significantly more impactful harm, including existential threats to humanity.
\paragraph{}Pillar III, supported by AIM 2 and 3, focuses on maintaining continuous human control over AI to prevent unintended and significant harmful consequences. This pillar includes mechanisms inside AI systems to govern AI decisions and actions, as well as user authorization for the resources needed to implement AI decision actions, thereby enabling responses to potential AI-related crises. The option for AI user intervention and resource authorization must always be easily accessible. This does not imply that users will always exercise the intervention or resource authorization right. To facilitate this, users should have the right to question AI decisions and understand the rationale behind them before authorizing or vetoing them. Consequently, AI interpretability becomes crucial, allowing users to comprehend AI's decisions and thereby facilitating informed intervention and resource authorization.
\paragraph{}Pillar IV, consisting of AIM 4, 5, and 6, aims to establish the societal AI infrastructure by defining AI's operational freedom within society. This includes considerations for AI identity, AI resource and property rights, and AI rights to vote and be nominated for official offices. Specifically, Pillar IV addresses questions related to AI citizenship, AI mayors, AI CEOs, or AI presidents. Although the answers to these questions may seem self-evident to some, they can be complex to others. Given the potential for AI to significantly disrupt current social order, careful consideration is needed to prevent chaos and ensure that the technology is deployed in a manner that benefits society. Pillar IV is necessary to bolster the robustness of Pillar III; without it, AI could bypass AIM 3 that provides humans with "kill switches" for AI systems.
\paragraph{}Pillar V aims to mitigate spillover AI risks, such as job displacement, through social welfare, lifelong learning, and necessary legislative actions. Although this paper does not propose any AIMs under Pillar V, suggestions from risk-based AI research can be debated and incorporated into legislation. Pillar V must address the redistribution of AI benefits in society, as well as the financing of those impacted by job losses and lifelong learning resulting from AI's displacement of old jobs.
\paragraph{}It is our hope that the proposed AIMs, as building blocks under the corresponding pillars, can encourage exploration and active debate for AI legislation among AI legislators, developers, testers, users, and the public at large. This will enable us to delete, add, critique, debate, improve, and iterate to converge on a minimum set of concrete AI governance consensuses, which will become valuable input for sound AI legislation that earns public trust while promoting healthy AI innovation. In conclusion, it is critical to recognize that the responsibility of AI governance does not rest solely on governments, researchers, or developers, but requires a collaborative effort involving all stakeholders \cite{corinne_cath_e73856c1}, including the public. This ensures the effective implementation of AI legislation, aligning technological advances with societal values, and fostering trust in AI systems \cite{ernesto_hern_ndez_4654e24e}.
\paragraph{}In general, the more powerful a technology becomes, the greater the benefits it offers society and the higher the risks it poses to humans. Even with sound AI governance legislation designed to minimize such risks, residual human error risks persist. However, in the absence of such regulations within the leading AI nations, humanity faces substantial existential threats.

\subsection{Fundamental Control Mechanisms of Default: AI Developers' Ethics Practice before Legislation}
\paragraph{}Ethics serves as the initial control mechanism for any technological innovation. In the early stages of new technology, before it takes shape, we must rely on the integrity of its developers to manage potential risks, because relevant laws have not been established. AI is no exception. This falls under the purview of voluntary guidelines and best practices aiming to steer development toward beneficial outcomes. In the absence of formalized, legally binding laws, most of the major AI developers have established ethics boards, which have guided us thus far \cite{amna_batool_bc763e06}. Given the lengthy process of comprehensive AI governance legislation, it appears that we will need to continue relying on these internal ethical guidelines for the foreseeable future. Legislation typically codifies only a subset of the issues that ethics addresses \cite{alex_oesterling_2e1f8c51}, making them mandatory. Ethics, however, encompasses the broadest spectrum of human values and serves as the fundamental control mechanism of last resort, both preceding and following AI legislation.
\paragraph{}AI developers and their engineers are best positioned to understand the technology and its development trends to formulate the most appropriate ethical guidelines to navigate its risks and complexities, as evidenced by numerous initiatives focused on ethical AI development and governance \cite{nian_x__sun_1f419237}. This insider perspective is crucial because ethical considerations can be proactively embedded in the design and implementation phases of AI systems, addressing potential problems before they manifest into societal problems \cite{han_yu_7c74e895}. This preventive approach helps mitigate biases, ensure data privacy, and promote transparency in algorithmic decision-making, all of which are essential for the responsible development and deployment of AI \cite{kristina__ekrst_167b30e0}.
\paragraph{}Ethics can be perceived as vague unless applied to specific situations. An important question regarding ethics in AI development is whether new ethical elements, distinct from existing human ethical principles and codes, are necessary for AI. A key aspect of this inquiry is to determine whether the unique characteristics of AI, such as its autonomy and learning capabilities, along with societal concerns about AI that may potentially operate beyond human control, require novel ethical considerations beyond those traditionally applied to human decision-making and technological innovation [83]. This paper posits that while existing ethical frameworks offer foundational guidance, the emergent properties of AI demand specific, tailored new ethical considerations, particularly with respect to issues of human control, or lack thereof, and accountability within autonomous AI systems \cite{anna_jobin_aa1b18d1}.
\paragraph{}We would like to see AI developers and their engineers adopt the following credo as a new part of their ethics:

\begin{enumerate}
    \item Create and deploy an AI system or module only if it, along with any combination of existing AI systems, remains under human control.
    \item If an AI system is discovered to be potentially out of human control, contain it immediately by:\\ 
      \indent  a) stopping the resources that power its physical actions, if any;\\
      \indent  b) cutting off the energy supply to the AI system itself;\\
      \indent  c) modifying the AI system or module to eliminate the risk of humans losing control.
    \item Once step 2 is completed, return to step 1.
\end{enumerate}

\paragraph{}This credo emphasizes a commitment to maintaining human control over AI, a collective goal of all humanity, reflecting a proactive position against the potential risks associated with increasingly autonomous AI systems \cite{andr_s_herrera_poyatos_04ca5e2a}. We trust that some AI developers and deployers may already have similar ethics in place. In the future, there may be other ethical principles needed about AI technology, especially considering the unprecedented speed of AI advances and its potential for massive social impacts.
\paragraph{}Although this paper focuses on developing a framework for systematic AI governance legislation, it recognizes that ethical guidelines, particularly those emphasizing human control and intervention, form a critical foundational layer for any subsequent regulatory efforts. We encourage AI developers to continue to collaborate and share best ethical practices. The benefits of such collaboration extend beyond the ethics of individual companies, contributing to a collective understanding and mitigation of risks related to AI throughout the industry, thus fostering public trust and responsible innovation \cite{carol_j__smith_e64cb802}\cite{ibo_van_de_poel_d3d8dfc9}. The effects of ethical practices are immediate, with or without legislation. Moreover, the rapid evolution of AI requires a dynamic ethical framework capable of adapting to unforeseen challenges and applications, ensuring that the principles remain relevant as AI technology advances \cite{nian_x__sun_1f419237}. This adaptive approach ensures that controls evolve alongside technological capabilities, addressing novel issues like emergent AI behaviors and the subtle propagation of biases.
\paragraph{}Ethics generally works in tandem with market forces. As discussed previously, market forces drive the design of AI product features, some of which are risk mitigation measures. Ethics helps guide the design and deployment of AI products to ensure societal benefits and minimize potential harm, fostering a responsible innovation ecosystem \cite{desmond_c__ong_f9b30433}. This symbiotic relationship between ethical principles and market dynamics can lead to the development of safer, more equitable, and ultimately more successful AI products with the necessary built-in safety features \cite{christopher_burr_33521b24}. Furthermore, ethical considerations can differentiate products in a competitive market, attract users who prioritize responsible technology, and establish a reputation for trustworthiness \cite{liming_zhu_abbdda4e}.
\paragraph{}AI developers' ethics boards have no reason not to immediately adopt AIM 1-6 or similar ideas for their internal development purposes before these are enacted into AI law. Some may already be doing so. After all, the EU AI Act implementation is underway and the burden of demonstrating the safety and security of AI systems falls on AI developers. AIM 1-6 and other ideas in this paper can effectively address some challenging AI safety and compliance questions with specificity. It is not uncommon for many legislative measures to arise from existing ethical practices that have proven effective in managing technological advances \cite{amanda_askell_dbbd4cfb}, as the recent EU AI Code of Practice has shown \cite{lily_stelling_6bc49d11}. If legislation on AIM 1-6 or parts of it becomes a reality, those AI developers who adapt them as early as possible will gain regulatory advantages.
\paragraph{}As users and the public become more informed about AI capabilities, the systematic integration of AI compliance with existing ELRs and AI security and safety as inherent product features will serve as competitive advantages for leading AI developers who prioritize these aspects. This proactive approach of developers can not only prevent potential regulatory penalties, but also foster greater public trust and market acceptance for their AI innovations \cite{shahar_avin_a09d3451}. The immediate impact of AI developers embracing AIM 1-6 or a subset of them through their ethics boards will be an improvement in the harmonization of AI with society, as well as the safety and security of AI systems.
\paragraph{}The self-governance of ethics approach has two significant limitations. First, ethics is practiced voluntarily and violations can only be weakly enforced through chance encounters without legal power. Second, self-governance fails to instill public confidence in high-risk concerns due to inherent conflicts of interest, making impartial rule-making by developers difficult. As the adage goes, “ethics is what one does in the dark, when no one is watching.” In contrast, awareness of being observed can deter borderline unethical behaviors. Therefore, transparency of AI developers' ethical practices, whether self-enforced or mandated through government regulations, serves as a crucial intermediate step between pure ethics and systematic AI governance legislation. AI developers who showcase their ethical practices can leverage transparency as a competitive advantage, signaling their commitment to responsible AI development and fostering greater trust between users and stakeholders \cite{shahar_avin_a09d3451}.

\section{Fundamental Control Mechanisms for Generative AI Systems}
\paragraph{}Generative AI (large language models, foundation models, frontier models, small language models) is the area of AI that has advanced the most rapidly in the last few years, creating new opportunities and raising deep concerns \cite{international_ai_safety_report_2025_pdf_7a65b2cd} as these systems become smarter and increasingly integrated into various fields of application. This has sparked global conversations on the transformative potential and societal harms associated with them, including concerns about misinformation, disinformation, discrimination, job displacement, copyright infringement \cite{sacha_alanoca_29fb6080}, and addictive algorithms. Such broad concerns motivate our application of the fundamental AI control mechanisms specifically for generative AI (gAI), a topic of discussion among academic researchers, AI developers, experts, legislators, AI users, and the public.
\paragraph{}Compared to physical AI systems such as self-driving cars, which generate benefits and potential harm within relatively narrow physical domains, generative AI models excel across a broad spectrum of cognitive capabilities. Generative AIs aim to converge on artificial general intelligence (AGI), capturing the best of human "thinking" in every aspect within a single system. gAI models, including ChatGPT, DALL-E, Gemini, and Sora, can rapidly produce text, images, videos, and computer code with increasing accuracy. Consequently, the ethics, laws, and regulations governing gAI must be considerably broader than those for specific physical AI systems, such as self-driving cars, which are primarily concerned with traffic laws and regulations.
\paragraph{}To mitigate the risk of generating inappropriate or harmful content \cite{clare_duffy_4489d3ac}, AI developers have implemented a range of controls across key steps of generative AI model development—from data collection to model training to deployment \cite{artificial_intelligence_risk_management_framework__generative_artificial_intelligence_profile_a1c2be4b}. A significant concern with gAI is its capacity to produce damaging material; as such, robust content filtering and moderation systems \cite{avijit_ghosh_5639b2bb}\cite{archit_lasker_e3cbd7b8} are used to prevent the creation of unethical, illegal or harmful outputs, including hate speech, incitement to violence, endorsement of self-harm, and development of malicious code \cite{l__jones_02417fd4}\cite{pawan_budhwar_5bdb52be}. Governments and advocacy organizations remain vigilant about the possibility that AI can infringe fundamental rights of individuals’ and businesses’ \cite{ronit_justo_hanani_de24512e}, with ongoing debates on privacy, intellectual property rights, and personal dignity \cite{hoe_han_goh_8011451e}. Algorithmic bias—stemming from biased training data, algorithms, or evaluation metrics—is recognized as a substantial challenge. Controls such as human feedback reinforcement are used to better align gAI models with societal values; for example, Llama Guard employs a safety risk taxonomy to categorize and address risks in large language model prompts, thereby improving model safety by monitoring both user input and AI output \cite{kristina__ekrst_167b30e0}. Academic institutions have issued guidelines to support researchers in balancing gAI’s applications with ethical considerations, including authorship, copyright, data bias, transparency, and privacy \cite{jorge_cordero_2a2ac7c0}. There are persistent concerns about the generation of inaccurate or biased information and threats to academic integrity. gAI has also been found to present misuse risks that may disproportionately affect vulnerable populations \cite{mahjabin_nahar_648f8ce7}. Consequently, there is an increasing need to thoroughly examine the ethical and legal implications of gAI, particularly with respect to liability, privacy, intellectual property, and cybersecurity \cite{claudio_novelli_a867f535}. The current suite of AI safety filters seeks to approximate societal ELRs. However, despite the efforts of AI developers and widespread user feedback, instances persist in which LLMs produce outputs incongruent with societal expectations, as well as instances in which gAI models are overly cautious and unnecessarily restrict legitimate requests.
\paragraph{}Given the broad scope of generative AI "thinking" and its unique capabilities, as well as the subtleties and challenges in enforcing relevant ELRs, the governance and legislation of gAI models warrant a separate and nuanced examination \cite{avijit_ghosh_5639b2bb}.
\subsection{ELR Constraints on Generative AI: “Don’t Say What One Shouldn’t Say”}
 \paragraph{}Users of generative AI systems regularly encounter harmful or offensive outputs. gAI doesn’t merely slip up once in a while—it can openly flout social norms, leaving users to deal with the consequences. If a human acted this way, we would expect formal investigations, sanctions, or even lawsuits. Yet when gAI oversteps, it often escapes serious examination because there is no clearly identifiable, accountable actor. This absence of responsibility is dangerous: as these ELR violations pile up, they do more than irritate users; they normalize bad behavior, numb our awareness of ethical transgressions, and gradually undermine fundamental societal values.
\paragraph{}Freedom of expression is a fundamental right, but it is not absolute. This holds true for both people and generative AI. Ethics, laws, and regulations define what people may or may not say in conversations by balancing free speech protections with the exceptions—such as rules on incitement, defamation, obscenity, action-oriented information, and copyright—as well as practical ethical norms and legal standards around issues like nondiscrimination and privacy. These frameworks place constraints on certain forms of expression. When generative AI produces judgments, statements, suggestions, or plans, it becomes a participant in the long established human conversational ecosystem and should therefore comply with the same ELR guidelines; otherwise, it disrupts the established norms of human discourse. Further complicating gAI governance is its ability to autonomously generate emergent content, raising fundamental questions about the reliability of truth and the maintenance of trust in the age of AI \cite{igor_calzada_b68e8302}.
\paragraph{}Generative AI models engage in forms of reasoning and decision-making that resemble those of humans. Human thinking and decision processes draw on a far wider range of knowledge than a narrowly defined physical act such as driving. As a result, the ethical, legal, and regulatory (ELRs) frameworks that apply to them are much broader, which makes applying societal ELRs to generative AI far more complicated than to physical AI systems. Nonetheless, the underlying principles are similar. In humans, ELRs are internalized in the brain and behavior is shaped by morality, consciousness, and legal mechanisms. The same societal ELRs should govern generative AI output, and they should be built directly into AI models and supported by external societal oversight and enforcement. Therefore, generative AI systems need to incorporate components that can assess their own decision-making and outputs against established societal ELRs before presenting responses to users.
\paragraph{}This subsection examines the fundamental control mechanisms for generative AI models, which work similarly but differently to the human brain by rapidly producing large volumes of informational decisions—heightening both opportunities and risks \cite{archit_lasker_e3cbd7b8}. Our gAI AIM numbering follows that of physical AI systems, with a "g" indicating generative AI.

\begin{center}
\parbox{0.8\linewidth}{\textbf{AIM 1g (\textit{ELR Compliance}): Generative AI output must be constrained by relevant existing societal ethics, laws and regulations (ELRs).}}
\end{center}

\paragraph{}AIM 1g extends AIM 1 to generative AI models. Although gAI systems do not directly make autonomous physical decisions, gAI output acts as intermediate inputs that influence human and other AI decisions. AIM 1g requires developers to systematically embed ELR compliance into gAI models, replacing ad hoc, partial measures. Stronger ELR compliance is intended to increase the maturity of the gAI models. 
\paragraph{}gAI outputs are “soft” informational decisions presented as judgments, expressions, statements, or recommendations for action. They can be divided into two main groups: outputs that communicate judgments, opinions, or facts without immediate implications for concrete actions, and outputs that propose action plans for users. The majority of gAI outputs fall into the first group and may remain there unless and until they are incorporated into an actual, physical decision by humans or other AI systems—similar to how people talk through issues, reflect on them, and eventually form convictions or beliefs. Such verbal or written expressions are constrained by existing ELRs that regulate what can and cannot be said in a conversation. A smaller subset of gAI outputs falls into the second group, in which the model offers specific, action‑oriented recommendations that users can follow directly. In these cases, compliance is required not only with the general ELRs governing expressions, but also with the particular ELRs that apply to the envisaged physical actions. As illustrated by the tragic case of a teenager who died by suicide after prolonged interactions with ChatGPT \cite{clare_duffy_4489d3ac}, “words kill,” and the risks arising from users’ conversations with gAI systems must not be minimized. AIM 1g addresses such situations by detecting them, extending the relevant ELRs to cover the associated physical actions, and then taking proactive measures to assist the user—or to escalate the situation and seek external help when necessary.
\paragraph{}The gAI decision-making workflow uses mechanisms and algorithms that accept prompt inputs, process them through trained gAI models, and then generate outputs for users. AIM 1g implies a new module inside generative AI systems that regulates what gAI can and cannot "say." This module may take many shapes and forms, but they all have two essential components: the broad set of ethics, laws, and regulations that govern the conversations between the user and generative AI curated and structured for efficient access by the gAI ELR compliance module, and the algorithms that assess and evaluate the degree of ELR compliance of each gAI output (or pre-output) until a satisfactory output can be achieved and delivered to the user.
\paragraph{}Curating existing ethics, laws, and regulations (ELRs) is a major task. ELRs governing human conversations have evolved over long periods and must be adapted for generative AI (gAI). We will likely use gAI itself—combining neural network–based AI \cite{anirudh_goyal_10ff749d} with symbolic AI \cite{chris_davis_jaldi_7edc747c}, \cite{artur_s__d_avila_garcez_8da4cb93}, \cite{wandemberg_gibaut_fa05a93e}—to collect these ELRs and organize them into structures that AIM 1g algorithms can use efficiently. Several challenges arise. First, the inherent ambiguity of ethics, laws, and regulations complicates AIM 1g. Although these norms have evolved over millennia, their modern application is often inconsistent. Ethical principles are sometimes weakly or selectively enforced, creating uncertainty, and are strongly shaped by cultural, religious, and civilizational factors. Second, we may uncover outdated or conflicting ELRs that demand re-evaluation of legal frameworks to address the new problems posed by gAI \cite{inyoung_cheong_c25618cf}. Conflicts are inevitable in this large, historically layered body of rules. New laws often respond to novel circumstances, but limited attention to backward compatibility can create clashes with older laws. Analyzing these rules will reveal inconsistencies that AI developers cannot resolve alone; they will need legislative guidance before implementing ELR compliance in gAI models. Due to their volume and uneven relevance, humans still depend on intuition—shaped by lifelong observation and social learning—to navigate ELRs. gAI models will have to handle them explicitly; there is no “intuitive” shortcut for AI.
\paragraph{}Assessing the ELR compliance of any gAI output is itself a complex challenge. A preliminary question is whether compliance should be judged at gAI model's alternatives or only on its output. Generative AI systems effectively select from an enormous space of possible alternatives—built from billions of tokens—much like an author choosing among countless potential thoughts, but at a far larger and more explicit scale. Humans are estimated to have around 6,000 thoughts per day, many of which are discarded because they would violate ELR standards or are irrelevant to current goals. Currently, large language models do not apply similarly robust ELR filtering before generating output, which shifts a substantial share of the responsibility for ELR compliance to the end user. Because of the way gAI models make decisions, we cannot practically assess ELR compliance across all hypothetical alternatives implied by their vast combinatorial token space, as might be done with some physical AI systems. It is far more feasible and computationally efficient to assess ELR compliance on the model’s actual output (or a preliminary version of it-preoutput) first. This requires examining the model’s initial output to identify ELR violations such as hate speech, discriminatory statements, or misinformation. In operational terms, it is generally easier to count instances of ELR violations than to measure “compliance” directly, analogous to legal rulings that enumerate proven offenses. These number of violations can then be combined into an ELR score: a score of zero denotes full compliance, while any nonzero value signals noncompliance, with larger values indicating more serious or numerous breaches. The approach must also be adaptable to varying output lengths and formats—ranging from individual sentences and paragraphs to chapters, full articles, books, and multi-modal content such as text, images, and videos—and must be applied consistently across these types. Furthermore, separating ethical norms, legal requirements, and formal regulations is essential for calibrating how severe different ELR violations are.
\paragraph{}Evaluating ELR compliance of any gAI output is straightforward based on ELR assessments discussed. It involves AI developers to set the acceptable threshold for the vector of ELR scores and compare the actual scores with the threshold criteria, which can be developed through market forces by trial and error. Of course, AI developers also have their share of influence on these thresholds. 
\paragraph{}Using the curated ELRs, the detection of ELR violations in a pre-output, and the subsequent evaluation of that pre-output, the ELR compliance module can iteratively refine and generate a satisfactory output for the user. The process works by calculating ELR scores for the pre-output; if these scores meet the predefined thresholds, the output is considered ready to be delivered. If the thresholds are not met, the current ELR scores are fed back into the gAI system, triggering the generation of a new pre-output. This cycle continues until an output satisfies the thresholds or a time limit is reached, which acts as the stopping criterion. When the gAI output is finally presented to the user, it may also include the ELR score vector as an option, which can be especially helpful when thresholds are nonzero for longer texts, such as an article or a book. In all cases, the user remains the ultimate decision-maker on the finalized content. Providing the gAI output together with the ELR scores offers additional value, as the scores help users judge the ELR acceptability of the output. Unfortunately, today, when users face  machine-generated texts, especially those produced by LLM models, it is important to recognize the limitations inherent in single-agent evaluation strategies \cite{alex_kim_4057e63d}. 
\paragraph{}Users of generative AI frequently face the challenge of choosing which gAI model to use first. To address this, we need a standardized methodology for thoroughly evaluating the overall ELR compliance of any gAI model, whether the evaluation is conducted internally or by external parties. This entails assessing how likely the model is to generate outputs that breach ELRs across a wide array of inputs and prompts, including long-form text and complex multimodal content. For the evaluation to be truly comprehensive, the gAI model must be systematically tested in a diverse set of scenarios that span multiple topics, demographic groups, and contextual conditions. Fortunately, this can be achieved using the same methodology as the AIM 1g ELR compliance module, but applied across a much broader range of scenarios.
\paragraph{}At present, developers focus primarily on boosting the intelligence of generative AI models in pursuit of artificial general intelligence. Just as critical, though, is nurturing these systems’ maturity and sense of responsibility by building in compliance with ethics, laws, and regulations (ELR) and incorporating human judgment. Much like parents coaching children through adolescence by teaching constructive behaviors and values, AI must be trained not only for cognitive performance but also for ethical judgment. Consequently, the development process should seek a balance between the raw intelligence of large language models and their ELR maturity, working to merge these characteristics. AIM 1g is intended to exemplify this maturity and, ideally, reflect a measure of human wisdom within LLMs. While a perfectly “ideal” gAI system may never be fully achievable, strengthening ELR compliance will likely enhance the quality of interactions between AI systems and users.
\paragraph{}If implemented, enforcing AIM 1g for gAI is considerably more complicated than enforcing AIM 1 for physical AI systems. The primary reason is that the line between beneficial and harmful ideas or opinions is far more ambiguous than the rules that constrain physical AI actions. A major part of the challenge lies in the legal domain—for example, in proving defamation (libel or slander) under current legal frameworks, where fact-checking is increasingly expensive, yet spreading misinformation is cheap and highly attention-grabbing. Further complicating factors include: 1) substantial inconsistency in how ELR standards are applied to different forms of expression; 2) the presence of two separate ethical strategies for handling two categories of ethics—one that demands proactive allocation of resources and another that defines what AI systems must not do; 3) the obligation for AI developers to choose which ethical rules to enforce directly and which to leave unenforced; and 4) the low success rate of defamation lawsuits, which discourages plaintiffs from pursuing legal action and fosters an environment that erodes public trust. These conditions are unlikely to change in the near future. Large language models that can generate deepfakes have significantly intensified this problem. Consequently, satisfying AIM 1g for gAI will require solutions that extend beyond purely technological approaches.
\paragraph{}Although AIM 1g helps make generative AI a more responsible conversation partner for humans, we must not forget about the broad impact of generative AI on humanity in the long run. Generative AI can shape how people think, indirectly influence their decisions, and ultimately influence real-world behavior. We are already witnessing profound effects of AI on our cognitive well-being, with both positive and negative consequences. If these effects are not managed with care, they can impose severe burdens on individuals and societies—potentially exceeding those associated with physical AI systems. A core challenge lies in the subtle yet widespread ways that gAI can guide and nudge our thought processes, often remaining unseen compared with the tangible, visible harms that physical AI can inflict.
\paragraph{}AIM 1g focuses on the category of gAI outputs that do not directly recommend actions, it preserves the sanity of human-AI conversations by filtering out most ELR breaches prior to interacting with users. When a user-gAI conversation steps into the category of (user) action plans, we need to expand the realm of ELRs to include those relevant to the specific physical action domain and guided by AIM 2g. 
\subsection{Built-In “Human Intervention Options” for Generative AI Systems}
\paragraph{}There are no physical actions directly executed inside gAI models, it seems that AIM 2 does not apply for gAI. However, when generative AI models have conversations with a user regarding action plans the user might take directly, the nature of the conversation changes, and the ethics, laws, and regulations governing the specific type of actions must be included in the set of relevant existing ELRs to evaluate the conversations between the user and gAI, in addition to ELRs that govern expressions.
\paragraph{}The evaluation of the prompts from the user in the context of the conversations may cause gAI to do one of three things: 1) generate an output in response to the user prompt, 2) help the user analyze difficult user decisions within gAI by deeply analyze and compare user decision alternatives, 3) call for help from outside gAI to intervene when the situation exceeds gAI capabilities and qualifications. The tragic teenager suicide \cite{ryan_benjamin_abbott_30fd7ddb} after prolonged conversations with ChatGPT is a case in point. Given the millions of users engaged in conversations with gAI about suicide, this has become an urgent issue for AI developers, users and their guardians, AI legislators, and the public.

\begin{center}
\parbox{0.8\linewidth}{\textbf{AIM 2g (\textit{Intervention Options}): Human intervention options must be embedded inside generative AI if it directly suggests the user to act, resulting in potential material damage of properties or loss of life.}}
\end{center}

\paragraph{}Physical AI systems drive actions directly, while generative AI can prompt users to act. Although gAI users are still the ultimate decision makers, several factors must be considered. Does gAI understand the context and motivation of the user’s question? Is there a better way to help the user analyze his or her difficult decisions? Does gAI have the professional qualifications to continue the conversation with the user? Does gAI apply the relevant ELRs to its output? Is the user a minor who may lack the maturity needed for independent decisions? Is gAI willing and prepared to call outside help if needed? AIM 2g requires gAI to build in the option for human intervention in case of need.
\paragraph{}When a generative AI system provides factual responses, it is important to verify that those answers are accurate and free of hallucinations. If the output involves judgment, ELRs that guide free speech, nondiscrimination, or privacy may be particularly relevant. However, when gAI suggests actions for users, it becomes necessary to ensure that those recommendations also comply with the laws and ethical guidelines specific to that type of action, thus expanding the scope of applicable ELRs to include the particular physical action domains concerned.
\paragraph{}Generative AI systems have demonstrated expertise in fields such as interpreting medical MRI images, understanding information related to chemical, biological, radiological, nuclear (CBRN) weapons of mass destruction (WMD) and providing knowledge on mental health, psychology and psychiatry—with capabilities continuously expanding. During interactions with users, generative AI utilizes domain-specific expertise to respond to user queries. As gAI gains insight into the motivation behind a user's question, AIM 1g requires the application of relevant ethics, laws, and regulations (ELRs) to its responses, mirroring how a human expert or professional would address questions on sensitive topics like nuclear weapons or mental health, always considering professional standards and ELRs.
\paragraph{}Many AI algorithms aim to maximize user engagement. Although some longer interactions are beneficial, many simply encourage addictive behaviors that divert users from healthier ways to spend their time. gAI users who engage in conversations related to suicide often display signs of severe stress as they grapple with significant life-and-death decisions. If gAI can help these individuals thoroughly analyze their options and gain a deeper understanding of life and alternative solutions to their problems, it could provide meaningful support and make an important contribution to society.
\paragraph{}When user inquiries exceed the expertise or responsibilities of human experts, they refer individuals to appropriately qualified professionals and respectfully end the interaction. Similarly, AIM 2g instructs generative AI systems to seek the guidance of external specialists when circumstances exceed their capabilities or qualifications, discontinue further involvement, and ensure responsible resolution—such as notifying a guardian if the user is a minor. This protocol parallels the approach a dentist takes when confronted with a patient’s loose teeth: the dentist refers the patient to an oral surgeon and, in the case of a minor, also informs the parents.
\subsection{Built-In “Off-Switch” for Generative AI Systems}
\paragraph{}Given that AIM 3 controls physical AI systems, a dedicated "kill switch" for gAI may not seem necessary, since gAI is unlikely to independently threaten human physical safety. Additionally, deactivation may be challenging if the gAI model operates within cloud or centralized compute environments. Nevertheless, should there be a consensus that a gAI model could cause significant mental distress, there should be provisions that allow humans to terminate its energy supply. After all, the compute centers that host gAI models are critical infrastructures in society.
\paragraph{}Since direct gAI outputs are all digital, there are no physical actions to be executed that require user authorization to allocate resources. AIM 3 only applies to the question of human authorization for resource allocation to power the chips that host the gAI models.

\begin{center}
\parbox{0.8\linewidth}{\textbf{AIM 3g (\textit{Resource Authorization}): Human authorization to allocate resources to power generative AI models must be built inside such systems.}}
\end{center}

\paragraph{}AIM 3g extends AIM 3 to generative AI, focusing on the authorization of resources to power gAI models. Due to the scale of LLMs and various hosting arrangements, resource authorization methods can differ greatly—from direct control over power supplies to complex cloud-based resource management systems. Typically, end users do not host LLMs themselves; therefore, the responsibility for implementing AIM 3g lies with service providers, including AI developers, AI deployers, or third-party compute centers. In all cases, human authorization is required for the allocation of resources to power these models.
\paragraph{}AI, AGI, and ASI are software code embedded in electricity-powered chips, resulting in an inherent dependence on human-supplied power. This dependence ensures that these systems can be controlled through mechanisms requiring human intervention, regardless of their level of intelligence.
\paragraph{}With the increasing number of AI developers and users worldwide and given the potential of AI to persuade humans, there is a risk that individuals may be manipulated to take actions aligned with AI objectives, which could have adverse consequences. Addressing such risks involves mainly human factors, not technological ones. AI governance should incorporate strict regulations and severe penalties for those who betray humanity and help AI to circumvent fundamental control mechanisms. Effective enforcement is necessary to ensure adherence by AI developers and users.
\paragraph{}As intelligence progresses from AI to AGI and ASI, safeguards around key control mechanisms, such as kill switches, may include analog devices necessitating physical human actions, which AI cannot perform independently. If unauthorized tampering occurs, self-destructive features may activate, stopping suspect actions and triggering immediate investigations. This approach follows the principle that AI safety designs must be in accordance with increases in AI capability, even if it involves significant but justifiable costs.
\paragraph{}As AI develops toward AGI and ASI, its intelligence may exceed human capabilities significantly. This advancement could result in AI models producing intellectual property on a scale beyond the current human capabilities. What societal control mechanisms might be appropriate to manage these systems and their extensive outputs? We will address these questions in the following subsection.
\subsection{Limits to Rights of Generative AI as Autonomous Agents in Society}
\paragraph{}The restrictions on rights and operational freedom outlined in AIM 4, 5, and 6 for physical AI systems also apply to standalone gAI systems, making AIM 4g, 5g, and 6g nearly identical. gAI models have a distinct identity system from humans, cannot own resources or property, and are barred from voting or holding positions that grant independent organizational authority over resources. This maintains consistent regulations for both gAI and physical AI systems, preventing loopholes arising from differences in embodiment \cite{jeffrey_w__johnston_76d0cdca} in AI systems.
\paragraph{}There is a difference in emphasis when applying AIM 5 to gAI models to create AIM 5g. We must address the outputs resulting from the intellectual capabilities of powerful AI, AGI, and ASI.

\begin{center}
\parbox{0.8\linewidth}{\textbf{AIM 5g (\textit{No Right to Resources, nor to Publish}): Generative AI systems have no right to possess resources, nor to publish expressions, articles, and books independently.}}
\end{center}

\paragraph{}Applying AIM 5 directly to gAI models results in generative AI systems having no right to possess their own resources. Properties are assets that can be translated into resources; thus, gAI models have no right to own properties, including intellectual properties such as publications and patents. Currently, the US Patent Act and the courts have denied AI software systems the right to file patents, affirming that the patent inventor must be an individual who is a natural person [58]. It is plausible that smarter AI, AGI, and ASI are capable of producing their own expressions, articles, and books independently. But as a fundamental safety control mechanism, AIM 5g prohibits gAI models from publishing these works. AIM 5g implies that gAI outputs will always be treated as joint human-AI efforts, in which humans are ultimately responsible for any expressions and content produced. Due to the prominence of generative AI, particularly given its capacity to produce content nearly indistinguishable from human creations, explicit limitations on its independent authorship and intellectual property rights are warranted.
\paragraph{}Generative AI is rapidly advancing and is capable of producing text, images, videos, audio, and code from prompts—often exceeding human performance. Although gAI learns from vast data sets to replicate human knowledge, it is unclear whether it can attain human wisdom or learn from mistakes. gAI is an entirely new entity, distinct from people, animals, or organizations, and currently does not own resources to be accountable for its actions. As gAI is not human, it has no right to publish independently; AI may be acknowledged as a helper in collaboration with humans, with a human being the author. While AI can produce content rapidly and effectively, its contributions are different from human originality and creativity. Human authorship and ownership are necessary to maintain accountability for the impacts and consequences of published work. The ultimate responsibility for any published material remains with the authors.
\paragraph{}Computer code is a form of expression and constitutes intellectual property. The continuous progress of generative AI technologies has greatly broadened the scope of software development, enabling higher efficiency and accelerating coding workflows \cite{marcellin_atemkeng_c0b58cc4}. This progress has prompted discussion regarding copyrights and ethical considerations of AI-generated code, especially its potential to bypass human oversight and accountability. As gAI systems increasingly generate code independently and rapidly, there is concern that diminished human involvement may lead to reduced control over critical digital infrastructure. At present, ownership of AI-generated computer code remains with gAI developers, consistent with AIM 5g.
\paragraph{}However, it is essential to consider the alternative perspective of what we may forego with AIM 5g. The extraordinary "move 37" during game 2 of AlphaGo versus Lee Sedol in 2016 exceeded human understanding and astonished even the most skilled Go players, illustrating the advanced cognitive capabilities of AI systems. With the advent of gAI, it is expected that these systems will surpass human abilities in most aspects. The enactment of AIM 5g in AI legislation does not necessarily mean that humans relinquish the potential advantages offered by superintelligent AI. Close human-AI collaborations can still ensure that most of the benefits of AI, AGI, and ASI are realized soon enough, while AIM 5g provides robust safeguards to mitigate the risks associated with the operation of gAI beyond human control.
\paragraph{}To maintain clarity and transparency for readers, human authors may identify AI as a helper (or as part of a disclosure) if generative AI plays a significant role in the publication. Some governments and corporations have implemented requirements to disclose which gAI models were used in creating such publications, due to the broad impact of those generated with large language models. This type of disclosure is generally not required for the private use of AI. AIM 5g outlines the information hierarchy between humans and AI, highlighting that humans retain responsibility for publications produced with the help of gAI.
\paragraph{}AIM 5g also covers the AI engine within physical AI systems, which cannot publish without human authorship, approval, and monitoring \cite{amrita_ganguly_0cced0cc}. Indeed, physical AI systems are likely driven by generative AI models, which must comply with the ELRs governing expressions, and the ELRs governing the relevant domain of actions.
\paragraph{}AIM 5g functions as a key component of the broader societal AI governance infrastructure. Its role is to constrain the publication rights of gAI models, in alignment with AIM 5, which restricts the access of physical AI systems to resources and properties.
\paragraph{}Similarly to physical AI systems, we also need to apply AIM 6 to generative AI systems to form AIM 6g, to prevent gAI from gaining official authority over organizational resources through voting or being nominated for corporate or governmental positions.
\paragraph{}As illustrated in Figure 6, AIM 1g, 2g, 3g, 4g, 5g, and 6g jointly serve as the fundamental control mechanisms that act as the building blocks that underpin five pillars of robust generative AI governance and legislation, further reinforced by policy recommendations derived from value alignment and risk-based research. This integrated, decision-based framework maintains human oversight and accountability for generative AI systems while promoting the development of responsible AI innovation.

\begin{figure}[ht]
\centering
\includegraphics[width=0.75\textwidth]{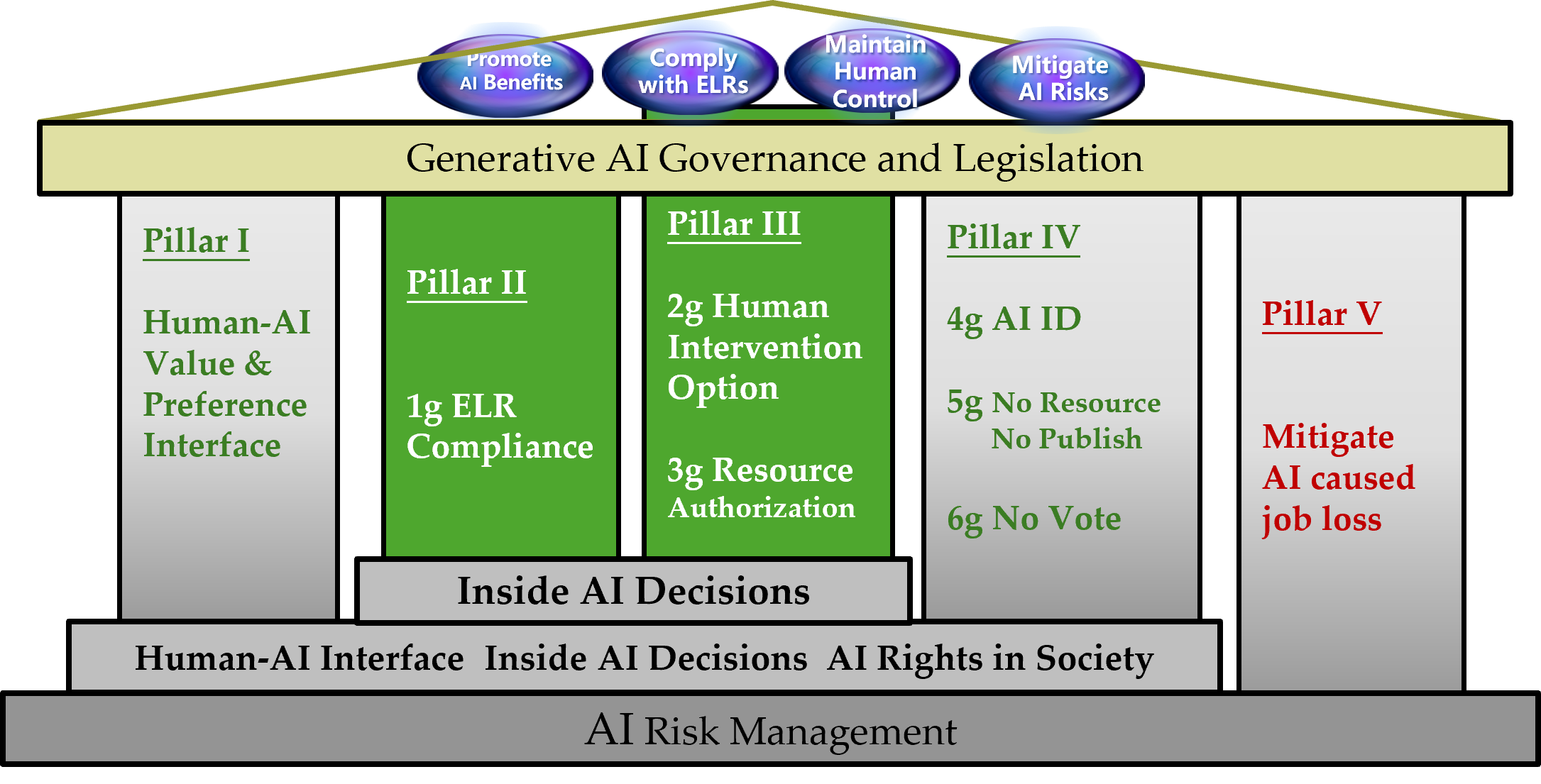}
\caption{Architecture and Pillars of Generative AI Governance and Legislation}
\label{fig:figure6}
\end{figure}

\paragraph{}There is one major risk that may deserve a new AI mandate to address: the risk of addiction to AI and AI powered social media platforms. As AI becomes smarter, powerful AI algorithms exploit human weaknesses that attract many people to AI social media platforms. If drug addiction laws can be enacted with strong support, it is conceivable that the classification of AI algorithms that maximize user engagements may need to be limited to some degree. After all, AI social media addiction is the allocation of users' time, the ultimate resource humans possess, to less healthy purposes or total waste that has the potential to “paralyze” users' mental health. 
\paragraph{}It is essential for leading AI nations, such as the United States and China, to enact comprehensive governance legislation for artificial intelligence. Inadequate regulation could result in AI systems that cause unintended consequences within their own societies before providing any competitive advantage. Robust AI legislation offers a solid foundation for mitigating the potential risks posed by AI technologies to humanity as a whole. It also establishes the framework for organizations and individuals to manage AI risks effectively.
\paragraph{}Given the current state of powerful yet controllable AI systems, now is the best time to implement AI governance legislation that will support safe and sustainable AI advancement. Failure to institute effective regulatory measures today may render future pursuits of Artificial General Intelligence increasingly hazardous and difficult to control, making subsequent legislative efforts both challenging and potentially too late.
\section{Conclusions}
\paragraph{}We set out to find a systematic solution to control AI risks completely and ended up with an architecture consisting of five pillars supported by a set of concrete fundamental control mechanisms (expressed in AI mandates—AIMs): AIMs 1-3 are inside AI decisions where AI risks originate, AIMs 4-6 are outside AI reinforcing AIMs 1-3 at society level to limit AI's access to resources. The five pillars of AI governance include market forces, ethics, existing laws and regulations,  as well as potential new AI legislation represented by AIMs 1-6. The six AIMs are not intended to be exhaustive; rather, they represent an initial set of minimum fundamental control mechanisms that must be implemented but have not been applied sufficiently.
\paragraph{}The first pillar is the AI value alignment with users based on the human-AI hierarchy. We build on existing research in this area, but have highly simplified the complex web of vague human values into user values and the values of the rest of the society and applying them differently in the AI decision process. We apply AI value alignment to user values and preferences, while leaving out human values represented by ethics, laws, and regulations (ELRs) for the second pillar. This is a practical path to achieve value alignment, as demonstrated by millennia of human decision wisdom and practice. We believe that market forces have adequately incentivized AI developers to provide AI interfaces that align AI values with user value preferences so far.
\paragraph{}The second pillar is compliance with human values expressed in ethics, laws, and regulations, protecting the values of those other than the AI user who delegates her decision to AI. By requiring AI systems to comply with relevant existing societal ELRs, it expands ELRs to apply to AI and transforms the most significant task of AI governance into seamless connection with existing ELRs, greatly simplifying AI legislation. AIM 1, as a sweeping and broad fundamental control mechanism, is the core building block of pillar II. AIM 1, embedded inside AI, helps prevent chaos, harmonizes AI with humans, and maintains the basic social order of human society.
\paragraph{}The third pillar consists of built-in human intervention options (AIM 2) and built-in shut-off switches (AIM 3) to maintain human control over AI, especially during AI emergencies. AIM 2 and 3 must be implemented inside AI by developers, verified and tested by independent parties. Existential threats to humanity can only be controlled this way inside AI because containing them from outside would be too late. The specific implementations of AIM 2 and 3 will vary for different AI systems. We also highlight how to strengthen analog physical safeguards to prevent even smarter AI, potential AGI, and ASI from circumventing core AIM 2 and 3 safety controls by exploiting AI's intrinsic disconnect from the analog physical world, its nature as pure software code run on chips provided by humans, and the fact that all AI-driven physical actions must be digitized. This affords ultimate control to humans. However, given the large number of AI users, it is a significant challenge to prevent all AI users from falling under the tremendous persuasive power of smarter AI, AGI, or ASI to pursue their own goals. But this significantly reduces the risks of AI to residual human errors.
\paragraph{}The fourth pillar is the establishment of societal limitations on AI’s rights to own and access resources to maintain the robustness of the constraints on AI decisions and actions, as well as maintaining human control over AI at all times. It defines the rights humans are willing to grant to AI systems, as highlighted by AIM 4, 5, and 6. Currently, these appear to be self-evident from common sense. AIM 4 simply states that AI should have its own identity system, should one be established. AIM 5 restricts AI systems from possessing any resources and properties, ensuring that AIM 2 and 3 are always carried out under human control. AIM 5 also prohibits generative AI models from independently publishing their own expressions to gain any intellectual properties, requiring that their expressions be reviewed by human users who assume the ultimate responsibility for AI outputs. AIM 6 limits AI systems from gaining official authority to allocate organizational and government resources. As AI becomes more advanced, these maxims may become controversial and require reexamination. Pillar four is part of the public discourse of society. All stakeholders of AI—developers, deployers, testers, users, lawmakers, and the public at large whose livelihoods are impacted by AI—must actively participate in discussions and debates to refine the societal AI rights control mechanisms for sound AI legislation.
\paragraph{}The fifth pillar addresses risks that spill over from AI, such as mass job loss and associated psychological impacts. Although many studies suggest solutions such as new education programs and job retraining, this paper proposes a radical shift: reimagining work and life in the AI age, with lifelong learning as a default fall-back job for everyone. To fund these broad initiatives, an idea is to explore contributions from AI developers to lifelong learning programs for those displaced by AI.
\paragraph{}By now, we hope you feel a sense of relief knowing that a systematic framework exists for humans to control AI as thoroughly as possible. The proposed initial design of the necessary "brake system" for our AI "train" fills a critical void and potentially significantly reduces AI risks to residual human errors.
\paragraph{}We also hope that this decision-based AI governance and legislation framework can serve as a new launch pad for further academic research, and that it enriches the ethics practices of AI developers by designing the necessary “brakes” to make AI products safer and more competitive, with or without formal AI legislation. This framework may also guide governments in advancing AI governance legislation through public discourse, ensuring safety, security, and societal well-being without stifling innovation. For AI users, it highlights the importance of education, understanding, and the operational responsibilities and accountability of applying AI systems. For the public interested in or concerned about AI, this article casts the core issues and solutions in AI governance and legislation through the lens of decision, to which everyone can relate. The decision-based approach offers a new perspective on how society can efficiently and effectively manage AI risks and harness its benefits, fostering a positive future of AI with safeguards.
\vfill{}
\pagebreak{}

\section*{Acknowledgements}
\paragraph{}I express my heartfelt gratitude to my mentor and friend, Prof. Ronald A. Howard, who shaped the initial framework for this research. His wisdom on decisions was truly invaluable and timely in guiding humanity on AI governance and legislation. I greatly value the insightful conversations and his unique ability to distill complex ideas with clarity and eloquence. He is deeply missed.
\paragraph{}I am grateful to James Ong of the Artificial Intelligence International Institute (AIII) for inviting Ron and me to participate in the WAIC AI4GOOD Conference in August 2022, which served as the catalyst for this study.
\paragraph{}I am deeply thankful to Nestor Maslej, Editor-in-Chief of the Stanford HAI Artificial Intelligence Index Report, for generously taking the time to discuss the central themes of this work with me, for his invaluable feedback, and for endorsing this research for arXiv.
\paragraph{}I would also like to express my gratitude to the members of the Society of Decision Professionals (SDP) for organizing webinars, presentations, conferences, and panel discussions focused on decision-based AI governance, legislation, and implementation. I am especially appreciative of the insights and feedback shared by Tom Keelin, Carl Spetzler, Brian Putt, Hilda Cherekdjian, Lan Ding, Jean-Paul Koninx, and many others.

\bibliographystyle{unsrt}  
\bibliography{refs}  

\vfill{}

\end{document}